%% file: main.tex
\documentclass[sigconf, nonacm]{acmart}
\usepackage[frozencache,cachedir=.]{minted}
\usepackage{balance}
\setminted{fontsize=\footnotesize}
\usepackage{listings}
\usepackage{booktabs}
\usepackage{multirow}
\usepackage{amsmath}
\usepackage{mathtools}
\usepackage{relsize}
\usepackage{stmaryrd}
\usepackage{xspace}
\usepackage{subcaption}
\usepackage{fancyvrb}
\usepackage{xcolor}
\usepackage{wrapfig}
\usepackage{tikz}
\usepackage[clock]{ifsym}
\usepackage{marvosym}
\usepackage{array}
\usepackage{graphicx}
\usepackage{amsthm}
\usepackage{lipsum}

\usetikzlibrary{positioning}
\tikzset{main node/.style={circle,fill=gray!20,draw,minimum size=0.45cm,inner sep=0pt}, }
\tikzset{invisible/.style={circle,fill=white!20,draw,minimum size=0.0cm,inner sep=0pt}, }
\usepackage{listings}
\usepackage{syntax}
\usepackage{mathpartir}
\usepackage{mdframed}
\usepackage{multicol}

\usepackage{algorithm}
\usepackage{algpseudocode}
\usepackage{xspace}
\newcommand\vldbdoi{10.14778/3712221.3712232}
\newcommand\vldbpages{651 - 665}
\newcommand\vldbvolume{18}
\newcommand\vldbissue{3}
\newcommand\vldbyear{2024}
\newcommand\vldbauthors{\authors}
\newcommand\vldbtitle{\shorttitle} 
\newcommand\vldbavailabilityurl{https://github.com/harp-lab/slog-lang1}
\newcommand\vldbpagestyle{empty} 
\newcommand{\slog}{\textsc{Slog}}

\newcommand{\bpra}{{BPRA}}

\newcommand{\revised}[1]{{#1}}
\newcommand{\dls}{\ensuremath{\mathcal{D\!L^{\exists!~\!}}}\!\xspace}

\newcommand{\ldls}{\ensuremath{\mathcal{D\!L_S}}\xspace}

\makeatletter
\newcommand*\bigcdot{\mathpalette\bigcdot@{.5}\ \!}
\newcommand*\bigcdot@[2]{\mathbin{\vcenter{\hbox{\scalebox{#2}{$\m@th#1\bullet$}}}}}
\makeatother

\newcommand{\cpp}{C\nolinebreak\hspace{0pt}\texttt{++}\:}
\usepackage{rotating}

\newcommand{\timeout}{{\small \Clocklogo{}}}

\newcommand{\core}{\dls}

\definecolor{lightgray}{rgb}{0.3, 0.3, 0.3} 
\definecolor{lightblue}{rgb}{0.85, 0.95, 1.0} 
\definecolor{lightgreen}{rgb}{0.9, 1.00, 0.9} 
\definecolor{midorange}{rgb}{0.6, 0.40, 0.2}
\definecolor{mgreen}{rgb}{0.7, 0.90, 0.85}
\definecolor{black}{rgb}{0.0, 0.0, 0.0}
\newcommand{\rtag}[1]{\textcolor{lightgray}{\bf{\texttt{#1}}}}

\newcommand{\comm}[1]{\textcolor{midorange}{#1}}

\newcommand{\HOLicon}[0]{\includegraphics[height=1em]{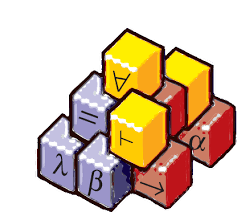}}
\newtheoremstyle{lemmaWithIcon}
  {3pt} 
  {3pt} 
  {\itshape} 
  {} 
  {\bfseries} 
  {.} 
  { } 
  {\thmname{#1}\thmnumber{ H#2}\thmnote{ (#3)}
   \HOLicon%
   } 
\theoremstyle{lemmaWithIcon}
\newtheorem{lemmah}{Lemma}

\newtheoremstyle{LemmaWithIcon}
  {3pt} 
  {3pt} 
  {\itshape} 
  {} 
  {\bfseries} 
  {.} 
  { } 
  {\thmname{#1}\thmnumber{ H#2}\thmnote{ (#3)}
   \includegraphics[height=1em]{figures/hol.png}%
   } 
\theoremstyle{LemmaWithIcon}
\newtheorem{theoremh}{Theorem}

\begin{document}
\title{Datalog with First-Class Facts}

\author{Thomas Gilray}
\affiliation{%
  \institution{Washington State University}
}
\email{thomas.gilray@wsu.edu}

\author{Arash Sahebolamri}
\affiliation{%
  \institution{Syracuse University}
}
\email{arash.sahebolamri@gmail.com}

\author{Yihao Sun}
\affiliation{%
  \institution{Syracuse University}
}
\email{ysun67@syr.edu}

\author{Sowmith Kunapaneni}
\affiliation{%
  \institution{Washington State University}
}
\email{sowmith.kunapaneni@wsu.edu}

\author{Sidharth Kumar}
\affiliation{%
  \institution{University of Illinois at Chicago}
}
\email{sidharth@uic.edu}

\author{Kristopher Micinski}
\affiliation{%
  \institution{Syracuse University}
}
\email{kkmicins@syr.edu}


\begin{abstract}

\revised{Datalog is a popular logic programming language for deductive reasoning tasks in a wide array of applications, including business analytics, program analysis, and ontological reasoning. However, Datalog's restriction to flat facts over atomic constants leads to challenges in working with tree-structured data, such as derivation trees or abstract syntax trees. To ameliorate Datalog's restrictions, popular extensions of Datalog support features such as existential quantification in rule heads (Datalog$^\pm$, Datalog$^\exists$) or algebraic data types (Souffl\'e). Unfortunately, these are imperfect solutions for reasoning over structured and recursive data types, with general existentials leading to complex implementations requiring unification, and ADTs unable to trigger rule evaluation and failing to support efficient indexing.}

\revised{We present \dls{}, a Datalog with first-class facts, wherein every fact is identified with a Skolem term unique to the fact. We show that this restriction offers an attractive price point for Datalog-based reasoning over tree-shaped data, demonstrating its application to databases, artificial intelligence, and programming languages. We implemented \dls{} as a system \slog{}, which leverages the uniqueness restriction of \dls{} to enable a communication-avoiding, massively-parallel implementation built on MPI. We show that \slog{} outperforms leading systems (Nemo, Vlog, RDFox, and Souffl\'e) on a variety of benchmarks, with the potential to scale to thousands of threads.}

\end{abstract}

\maketitle

\pagestyle{\vldbpagestyle}
\begingroup\small\noindent\raggedright\textbf{PVLDB Reference Format:}\\
\vldbauthors. \vldbtitle. PVLDB, \vldbvolume(\vldbissue): \vldbpages, \vldbyear.\\
\href{https://doi.org/\vldbdoi}{doi:\vldbdoi}
\endgroup
\begingroup
\renewcommand\thefootnote{}\footnote{\noindent
This work is licensed under the Creative Commons BY-NC-ND 4.0 International License. Visit \url{https://creativecommons.org/licenses/by-nc-nd/4.0/} to view a copy of this license. For any use beyond those covered by this license, obtain permission by emailing \href{mailto:info@vldb.org}{info@vldb.org}. Copyright is held by the owner/author(s). Publication rights licensed to the VLDB Endowment. \\
\raggedright Proceedings of the VLDB Endowment, Vol. \vldbvolume, No. \vldbissue\ %
ISSN 2150-8097. \\
\href{https://doi.org/\vldbdoi}{doi:\vldbdoi} \\
}\addtocounter{footnote}{-1}\endgroup

\ifdefempty{\vldbavailabilityurl}{}{
\vspace{.3cm}
\begingroup\small\noindent\raggedright\textbf{PVLDB Artifact Availability:}\\
The source code, data, and/or other artifacts have been made available at \url{\vldbavailabilityurl}.
\endgroup
}

\section{Introduction}
\label{sec:intro}

\revised{Datalog has emerged as a de-facto standard for iterative reasoning applications such as business analytics~\cite{datomic, aref2015design}, context-sensitive program analysis~\cite{bravenboer2009doop,antoniadis2017porting}, and ontological query answering~\cite{motik2019maintenance}. Datalog's success has been a harmony between the power of recursive queries and increasingly high-performance implementations, effectively leveraging parllelism on multiple cores~\cite{jordan2016souffle,byods}, clusters~\cite{shkapsky2016big,socialite}, or even GPUs~\cite{sun2023gdlog}. Modern implementations are efficiently operationalized via compilation to iterated joins over efficient data structures (e.g., tries~\cite{jordan2019specialized}, BTrees~\cite{jordan2019brie}, or hash tables~\cite{kumar:hipc:2019}) representing sets of tuples over word-sized values.}

\revised{Unfortunately, while Datalog's syntactic restrictions lead to straightforward decidability results, those same restrictions make Datalog ill-suited to applications which reason inductively over tree-shaped data such as derivations or syntax. In this work, we introduce \dls{}, an extension of Datalog to enable \emph{first-class} facts. \dls{} takes inspiration from languages such as Datalog$^\pm$ and Datalog$^\exists$, which allow existential quantifiers in the head of rules~\cite{cali2009general}. \dls{} restricts Datalog$^\exists$ to a \emph{uniqueness quantification} where every fact implies the existence of its own identity. Such a choice meaningfully enhances Datalog's expressivity---enabling a form of equality-generating dependency~\cite{Bellomarini:22} over tree-shaped facts. In \dls{}, every tuple is uniquely identified by a Skolem term unique to the fact. Such a choice rapidly enables computing over tree-shaped data, while yielding a lock-free, communication-avoiding implementation approach, compatible with state-of-the-art parallelizing implementation techniques for Datalog~\cite{cc,kumar:2020}.}

\revised{We formalize \dls{} as the syntactic extension of Datalog to include \emph{unique existential quantification} ($\exists!$) in the head of rules. This restriction ensures that every fact is identified by a Skolem term unique to the fact. We present a chase-based semantics for \dls{}, whose implementation is significantly simplified by the restriction that all facts are structurally unique, sidestepping the challenges of unification by associating each (sub)-fact with a unique intern id. The syntactic restrictions of \dls{} forbid unifying the head with the body, thereby preventing cyclic reference and yielding finite expansion sets. While \dls{} is only semi-decidable in general, running subsequent iterations of a \dls{} query will always produce increasingly-larger facts, with each iteration adding finitely-many facts. Practically, \dls{}'s properties fortuitously enable a trivially-parallel implementation as a simple extension of modern parallel relational algebra implementations~\cite{kumar:hipc:2019,cc,isc-kumar}. }

\revised{Specifically, our contributions are as follows:}

\revised{(1) We introduce \dls{}, a Datalog with first-class facts, i.e., in which all facts are uniquely identified via a nested Skolem term. We present a semantics for \dls{} based on the restricted chase, and then present \ldls{}, a language which compiles down to \dls{}; \ldls{} enables more natural programming with directly-nested facts, equivalent in power, and is the basis for our implementation.} 

\revised{(2) We present a series of applications demonstrating \dls{}'s generality and relevance to a broad array of fields. Specifically, we discuss \dls{}'s application to provenance, algebraic data (\emph{e}.\emph{g}., as included in Souffl\'e), functional programming (as in IncA~\cite{pacak2022functional}), structural abstract interpretation~\cite{VanHorn:2010}, and type systems~\cite{pacak:20,tapl}. }


\revised{(3) We implement \slog{}, a fully-featured, data-parallel engine, which compiles \ldls{}, with syntactic sugar, to a Message Passing Interface (MPI) based runtime. While there are no explicit constructs for data parallelism in \ldls{}, the restrictions of \dls{} enable leveraging recent work in (balanced) parallel relational algebra, \bpra{}, which uses hash-based distribution strategies and load-balancing, scaling to thousands of threads. We show that \bpra{} can be extended to support \dls{} with just a small change to its implementation: a trivially-parallelizable mechanism for tuple deduplication, to intern and assign unique references for new tuples (\S\ref{sec:slog}).}

\revised{(4) We evaluate \slog{} on a series of applications in graph analysis with provenance and program analysis, both with and without materializing provenance information, illustrating how our framework supports a unified perspective on first-class facts as provenance, at most, or as unique enumerated references, at least. 
Operationally, our approach enables us to design rich, highly-parallel static analyses which leverage \dls{} as a foundation.}
\revised{We evaluate \slog{} by implementing these applications and performing a series of experiments showing scaling and comparisons against other state-of-art systems. These include (a) why-provenance comparisons versus Nemo~\cite{ivliev2023nemo}, VLog~\cite{urbani2016column}, and RDFox~\cite{rdfox}, (b) context-sensitive analysis of Linux, and (c) results showing an exponential gap versus Souffl\'e when using algebraic data types (\S\ref{sec:eval}).}


\section{Datalog with First-Class Facts}
\label{sec:first-class-facts}

\revised{Positive Datalog (\ensuremath{\mathcal{D\!L^+}}) rules contain a head and body clauses:
\[
H(x, y, \ldots) \leftarrow B(a, b, \ldots) \wedge \ldots
\]
In \ensuremath{\mathcal{D\!L^+}}, the body contains only positive clauses, and a rule is \emph{safe} whenever every variable in the head ($x, y, \ldots$) is contained in the body; such a restriction ensures finite solutions. It is trivial to extend the definition of safety to stratified negation (i.e., \ensuremath{\mathcal{D\!L}}): every variable that occurs in a negated subgoal must also appear in a positive subgoal. Such safety conditions ensure that \ensuremath{\mathcal{D\!L}} programs may be evaluated in a bottom-up manner, iterating an immediate consequence operator to a necessarily-finite fixed point.}

\revised{Despite being a popular implementation platform for an array of applications, Datalog's restriction to flat rules over atomic constants has proven limiting in applications such as ontological reasoning. Such settings necessitate structurally representing and computing over knowledge, well beyond the power of Datalog's flat rules. In a response to this limitation, a growing array of languages build upon \ensuremath{\mathcal{D\!L}} to enable reasoning over ontologies, including \ensuremath{\text{Datalog}^\pm} and \ensuremath{\text{Datalog}^\exists}. In parallel, a growing breadth of work in programming languages compiles higher-order functional programs~\cite{pacak2022functional} and type systems~\cite{pacak:20} to Datalog, grappling with its semantic limitations via monomorphization or ad-hoc extensions.}

\subsection{\core{}: its syntax and chase semantics}
\label{sec:dlexistsbang}

\revised{The core of our contribution is to identify a price point for the implementation of high-performance, parallel Datalog engines which also effectively enables applications such as provenance, ontology reasoning, structural type systems, and functional programming. Specifically, we introduce \dls{}, an extension of Datalog to enable first-class facts. \dls{} introduces a single syntactic extension to Datalog: every fact is assigned an \emph{identity}, which may be otherwise used as any other base value: referenced (selected and joined upon) in any body clauses and used to populate any (non-\textit{id}) column of the head clause. Rules in \dls{} take the form:
\[
 \big(\exists!\textit{id}_H.\ \textit{id}_H = H(x, y, \ldots)\big)
 \leftarrow
id_B =  B(a, b, \ldots) \wedge \ldots
\]
The $\exists!$ in \dls{} refers to the fact that the identity of the deduced fact is \emph{uniquely} determined for this rule and all of the values used in the head.  Unlike \ensuremath{\text{Datalog}^\pm} and \ensuremath{\text{Datalog}^\exists}, the logical variable quantified by $\exists$ does not appear as a column but instead serves as an annotation associated with the generated tuples. This design choice means that the identity of the head clause cannot be unified with logical variables in the body clauses or with other logical variables in the head. For example, the following two rules are not valid \dls{} queries:
\begin{example}
Ill-formed unification of annotations in \dls{}:
\[
\begin{array}{lcl}
     c = H(a, b) & \leftarrow & \textit{id} = B(a, b, c) \\ 
     \exists! c. H_1(a, b) \wedge H_2(a, c)  & \leftarrow  & B(a, b)
\end{array}
\]
\end{example}
By excluding unification between the identity of the head clause and logical variables in the body or head, we ensure that identity values are used in a strictly acyclic manner. This restriction prevents undesired non-termination issues that can arise from chasing cyclic dependencies. Meanwhile, forbidding identity unification across head clauses enables more parallel implementations of the \dls{} query engine. It simplifies the satisfaction check for existentially quantified queries to merely verifying whether the inferred head clause tuple already exists, thereby avoiding extra communication in distributed settings. We will elaborate on this in Section~\ref{sec:mpi}.}

\revised{The use of the existential quantifier in rule heads extends Datalog with \emph{tuple-generating dependencies} (TGDs), which trigger tuple generation based upon the insertion of other tuples in the database. While such an extension significantly enhances Datalog's expressivity, it comes at a cost: adding TGDs makes queries undecidable in general and leads to infinite results. As a response, there has been a significant amount of interest in tractable restrictions to TGDs which ensure decidability~\cite{ijcai2017p128,leone2012efficiently,cali2009general}. Much of this work leverages the \emph{restricted chase}~\cite{carral2017restricted}, a sound and complete algorithm for query answering over ontologies of disjunctive existential rules. The restricted chase discovers all derivations in bottom-up fashion, generating labeled-nulls and unifying Skolem terms on demand.}

\begin{algorithm}

\caption{\revised{applyRules($\Sigma$, $\Delta$, $i$)}}
\label{alg:chase}
\begin{algorithmic}[1]
\State $\Delta^{i+1} = \emptyset$
\ForAll{$\big(\exists! \textit{id}_H. H(x,...) \leftarrow \phi\big) \in \Sigma$}
    \ForAll{match $\theta$ of $\phi$ over $\Delta^{[0,i]}$ with $\phi\theta \cap \Delta^i \neq \emptyset$}
        \If{$\Delta^{[0,i]} \not\models  \big(\exists! \textit{id}_H. \textit{id}_H = H(x,y,...)\big) \theta$}
            \State $\theta' = \theta \cup \{\textit{id}_H \mapsto n\}$ where $n$ is a fresh null value
            \State $\Delta^{i+1} = (\Delta^{i+1} \cup \big( \textit{id}_H = H(x,...)\big) \theta') \setminus \Delta^{[0,i]}$
        \EndIf
    \EndFor
\EndFor
\State $i = i + 1$
\end{algorithmic}

\end{algorithm}

\revised{We give the semantics of \dls{} via a variant of the restricted chase. The function $\textit{chase}(\Sigma, \mathcal{I})$ iteratively builds a frontier of newly-generated facts in $\Delta$, a (potentially infinite) stream of facts; the iteration, $i$, is initially set to $0$ and $\Delta^0$ is set to $\mathcal{I}$, the input database. Algorithm~\ref{alg:chase} is then repeated until $\Delta^{i-1} = \emptyset$. The algorithm finds all possible matches $\theta$ in the current generation, and then checks whether a first-class fact $H(x,...)$ exists in the database $\Delta^{[0,i]}$, substituting the head via the match $\theta$. If not, the algorithm generates a fresh labeled null, building the updated substitution $\theta'$;  last, $\theta'$ is applied to the rule's head, $\big( \textit{id}_H = H(x,...)\big)$, to materialize the new first-class fact. Across iterations, $\Delta$ records increasingly-tall facts, equating each fresh null with a novel Skolem term.
\begin{example}
Consider the following two \dls{} rules:
\[
\begin{array}{rcl}
 \exists ! \textit{id}_T .\ \textit{id}_T = T(\textit{id}_G)  &\!\!\!\leftarrow\!\!\! & \textit{id}_G = G(x) \land x \neq A() 
\\
 \exists!\textit{id}_{T'}.\ \textit{id}_{T'} = T(\textit{id}_{G'})  &\!\!\!\leftarrow\!\!\!& \textit{id}_T = T(\textit{id}_G) 
\\&& \land \textit{id}_G = G(\textit{id}_{G'}) \land \textit{id}_{G'} = G(x) \\
\end{array}
\]
\end{example}
Consider we start with $\mathcal{I}$ such that:
\[
\begin{array}{rl}
\Delta^0 = \{&\!\!\!\!\!\textit{id}_A = A(),
\textit{id}_{G1} = G(\textit{id}_A = A()) \\
&\!\!\!\!\!\!,\textit{id}_{G2} = G\big(\textit{id}_{G1} = G(\textit{id}_A = A())\big) \}
\end{array}
 \]
Notice that the database is referentially closed in the sense that every referenced sub-fact is included in $\Delta^0$; we will more rigorously define this property shortly. Continuing the chase, we conclude:
\[
\begin{array}{rcl}
\Delta^1 & = &\{\textit{id}_{T2} = T(\textit{id}_{G2} =
 G(\textit{id}_{G1} = G(\textit{id}_A = A())))\}\\
\Delta^2 & = &\{\textit{id}_{T1} = T(\textit{id}_{G1} = G(\textit{id}_A = A()))\}\\
\Delta^3 & = & \emptyset \\
\end{array}
 \]}



\revised{Like Datalog$^\pm$ and Datalog$^\exists$, the restricted chase for \core{} is undecidable in general, evidenced by the fact that it is possible to build an interpreter for the $\lambda$ calculus in \ldls{} (Figure~\ref{fig:lambdacalc}). However, the behavior of the restricted chase given in Algorithm~\ref{alg:chase} is simplified by the fact that facts are uniquely identified. This fact eliminates the challenge of unification, as facts are simply assigned a primary key associated with a Skolem term. With respect to chase termination, the programs for which Algorithm~\ref{alg:chase} does not terminate are precisely those which generate an unbounded number of \textit{id}s. For any \dls{} program which produces facts bounded by any finite $n$, Algorithm~\ref{alg:chase} necessarily terminates; in \S~\ref{sec:apps-dl}, we use this fact to establish the decidability of the Datalog subset of \core{}.}

\subsection{From \core to \ldls{}}
\label{sec:core}

%
\begin{figure}[h!]
\resizebox{6.25cm}{!}{$
\begin{array}{rcl}
  \langle \textit{Prog} \rangle  & ::= &\langle \textit{Rule}\rangle^\ast \\
  \langle \textit{Rule}\rangle  & ::= &\textit{R}(\langle \textit{Subcl}\rangle, \ldots)
 \leftarrow \langle \textit{Clause}\rangle, \ldots\\
  \langle \textit{Clause}\rangle &::= &\textit{id} = \textit{R}(\langle \textit{Subcl}\rangle, \ldots)\\
  \langle \textit{Subcl}\rangle & ::=& \textit{R}(\langle \textit{Subcl}\rangle, \ldots) \mid \langle \textit{Var}\rangle \mid \langle \textit{Lit}\rangle\\
  \langle\textit{Lit}\rangle   & ::= &\langle\textit{Number}\rangle \mid \langle\textit{String}\rangle \mid ... \\
\end{array}$
}
\vspace{-0.25cm}\caption{Syntax of \ldls{}: \textit{R} is a relation name.}
\label{fig:dls-syntax}
\vspace{-0.2cm}
\end{figure}

\revised{We now extend \core{} to \ldls{}, syntactic sugar on top of \core{} which allows syntactically-nested facts and patterns, such as $G(G(x))$. 
The syntax of \ldls{} is shown in Figure~\ref{fig:dls-syntax}. As in
Datalog, a \ldls{} program is a collection of Horn clauses. Each
rule $R$ contains a set of body clauses and a head clause, denoted by
$\textit{Body}(R)$ and $\textit{Head}(R)$ respectively. \ldls{} programs must also be well-scoped: variables appearing in a
head clause must also be contained in the body; notice that \ldls{} omits the existential quantification and assignment in the head, making the quantifier implicit. The definition of \ldls{} is given via a syntax-directed translation given in Figure~\ref{fig:compiling}, which traverses rule bodies top-down to flatten their structure, decomposing nested clauses and flattening them to expose explicit identities. For each subclause, the \textit{subcl} function is used to recursively flatten it to a set of flat clauses and a single value that uniquely identifies the original subclause without containing any structured subclauses. }

\begin{figure}
\begin{align*}
  \textit{flatten} :\ 
  & 
  \langle\textit{Rule}\rangle \rightarrow \langle\textit{Rule}\rangle
  \\
  \textit{flatten}(rule) \triangleq\ 
  & 
  \textit{Head}(rule) \leftarrow \!\!\!\!\!\!\!\!\!\bigcup_{\textit{cl} \in \textit{Body}(\textit{rule})}\!\!\!\!\!\!\!\!\!\! \textit{clause}(\textit{cl})
  \\
  \textit{clause} :\ 
  & 
  \langle\textit{Clause}\rangle \rightarrow \mathcal{P}(\langle\textit{Clause}\rangle)
  \\
  \textit{clause}(\textit{id} = \textit{R}(\ldots,\textit{scl}_j)) \triangleq\ 
  & 
  \{\textit{id} = \textit{R}(\ldots, \textit{scl}'_i, \ldots)\} \cup \bigcup_{i} \textit{cset}_i
  \\
  \textrm{where} &\ 
  \textit{scl}'_i, \textit{cset}_i = \textit{subcl}(\textit{scl}_i)
  \\
  \textit{subcl} :\ 
  & 
  \langle\textit{Subcl}\rangle \rightarrow \langle\textit{Subcl}\rangle \times \mathcal{P}(\langle\textit{Clause}\rangle)
  \\
  \textit{subcl}(\textit{R}(\ldots,\textit{scl}_j)) \triangleq\ 
  & 
  \textit{id}, \{\textit{id} = \textit{R}(\ldots, \textit{scl}'_i, \ldots)\} \cup \bigcup_{i} \textit{cset}_i
  \\
  \textrm{where} &\ 
  \textit{scl}'_i, \textit{cset}_i = \textit{subcl}(\textit{scl}_i)
  \text{ and } \textit{id} \text{ is fresh}
  \\
  \textit{subcl}(\textit{scl}) \triangleq\ 
  & 
  \textit{scl}, \{\ \!\}\ \textrm{where}\ \textit{scl} \in \langle\textit{Var}\rangle\cup\langle\textit{Lit}\rangle
\end{align*}
\caption{\revised{Compiling \ldls{} into \core{}}}
\label{fig:compiling}
\end{figure}


\subsection{Fixed-point Semantics}

\revised{The fixed-point semantics of a \ldls{} program $P$ is given via the
least fixed point of an \emph{immediate consequence} operator
$\textit{IC}_P : \textit{DB} \rightarrow \textit{DB}$. Intuitively,
this immediate consequence operator derives all of the immediate
implications of the set of rules in $P$. A
database $\textit{db}$ is a set of facts ($\textit{db} \in \textit{DB}
= \mathcal{P}(\textit{Fact})$). A fact is a clause without variables:
\begin{align*}
\textit{Fact} &::= \textit{R}(\textit{Val}, \ldots) & \quad
\textit{Val}  &::= \textit{R}(\textit{Val}, \ldots )\ \mid \textit{Lit}
\end{align*}
In Datalog, $\textit{Val}$s are restricted to a finite set of atoms
($\textit{Val}_{\textit{DL}} ::= \textit{Lit}$). To define $IC_P$, we
first define the immediate consequence of a rule $IC_R : DB
\rightarrow DB$, which supplements the provided database with all the
facts that can be derived directly from the rule given the available
facts in the database:}
\begin{align*}    
\textit{IC}_R(db) \triangleq\ & db\ \cup \bigcup \bigl\{ \textit{subfact}(\textit{Head}(R)\big[\overrightarrow{v_i\backslash x_i}\big]) | \\
& \{\overrightarrow{x_i \rightarrow v_i}\} \subseteq (\textit{Var} \times \textit{Val})\ \wedge
\textit{Body}(R)\big[\overrightarrow{v_i \backslash x_i}\big] \subseteq db \bigr\}
\end{align*}

The \textit{subfact} function has the following definition:


\begin{centering}
  \[
  \begin{array}{rcl}
  \textit{subfact}\bigl(\textit{R}(\textit{item}_1,\ ...\ ,\textit{item}_n)\bigr) &\triangleq& 
  \{\textit{R}(\textit{item}_1,\ ...\ , \textit{item}_n)\} \\ 
  & & \cup \bigcup_{i \in {1 ... n}} \textit{subfact}(\textit{item}_i)   \\
  \textit{subfact}(v)_{v \in \textit{Lit}} &\triangleq& \{\}
  \end{array}
  \]
\end{centering}

\revised{The purpose of the \textit{subfact} function is to ensure that all nested facts are included in the database, a property we call \emph{subfact-closure}.
This is also the semantic counterpart to \textit{flatten} from Figure~\ref{fig:compiling}.
The immediate consequence of a program is the union of the immediate
consequence of each of its constituent rules, $
\textit{IC}_P(\textit{db}) \triangleq \textit{db} \cup \bigcup_{R \in
  P} \textit{IC}_R(\textit{db})$. Observe that $\textit{IC}_P$ is
monotonic over the lattice of databases whose bottom element is
the empty database. Therefore, if $\textit{IC}_P$ has any fixed
points, it also has a least fixed point \cite{tarski1955lattice}.
Iterating to this least fixed point directly gives us a na\"ive, incomputable
fixed-point semantics for \core{} programs.
Unlike pure Datalog, existence of a finite fixed point is not guaranteed in
\core{} This is indeed a reflection of the fact that \core{} is
Turing-complete. The \core{} programs whose immediate consequence
operators have no finite fixed points are non-terminating.}
Results formalized in Isabelle/HOL proof assistant are marked with \HOLicon.

\begin{lemmah}
The least fixed point of $\textit{IC}_P$ is subfact-closed. 
\end{lemmah}

It is worth pointing out that the fixed point semantics of Datalog is similar, the only difference being that the $\textit{subfact}$ function is not required, as Datalog clauses do not contain subclauses.

\subsection{Model-theoretic Semantics}

\revised{The model-theoretic semantics of \core{} closely follows the model
theoretic semantics of Datalog, as presented in, e.g.,
\cite{ceri1989you-datalog}.} The \emph{Herbrand universe} of a \core{}
program is the set of all of the facts that can be constructed from
the relation symbols appearing in the program. Because \core{} facts
can be nested, the Herbrand universe of any nontrivial \core{} program
is infinite. For example, $\mathbb{N}$ may be encoded in
\core{} using the zero-arity relation \lstinline{Zero} and a unary
relation \lstinline{Succ}. The Herbrand universe produced by just
these two relations, one zero arity and one unary, is inductively infinite.

A \emph{Herbrand Interpretation} of a \core{} program is any subset of its Herbrand universe that is subfact-closed. I.e., if $I$ is a Herbrand Interpretation, then $I = \bigcup \{\textit{subfact}(f) |\ f \in I\}$. For Datalog, the Herbrand Interpretation is defined similarly, with the difference that subfact-closure is not a requirement for Datalog, as Datalog facts do not contain subfacts.

Given a Herbrand Interpretation $I$ of a \core{} program $P$, and a rule $R$ in $P$, we say that $R$ is true in $I$ ($I \models R$) iff for every substitution of variables in $R$ with facts in $I$, if all the body clauses with those substitutions are in $I$, so is the head clause of $R$ with the same substitutions of variables.
\begin{align*}  
& I \models R\ ~\textrm{iff} ~
 \forall \{\overrightarrow{x_i \rightarrow v_i}\}\ .\ \textit{Body}(R)\big[\overrightarrow{v_i\backslash x_i}\big] \subseteq I \longrightarrow \textit{Head}(R)\big[\overrightarrow{v_i\backslash x_i}\big] \in I 
\end{align*}

If every rule in $P$ is true in $I$, then $I$ is a \emph{Herbrand model} for $P$. The denotation of $P$ is the intersection of all Herbrand models of $P$. We define $\mathbf{M}(P)$ to be the set of all Herbrand models of $P$, and $D(P)$ to be $P$'s denotation. Then, $D(P) \triangleq \!\!\!\!\!\!\bigcap\limits_{I\in\ \mathbf{M}(P)}\!\!\!\!\! I$. Such an intersection is also a Herbrand model:

\begin{lemmah} The intersection of a set of Herbrand models is also a Herbrand model.
\end{lemmah}

Unlike Datalog, nontrivial \core{} programs have Herbrand universes
that are infinite. Thus, a \core{} program may have only infinite
Herbrand models. If a \core{} program has no finite Herbrand models,
its denotation is infinite and so no fixed-point may be finitely
calculated using the fixed-point semantics. We now relate the operational
semantics of \core{} to its model-theoretic semantics: we elide a detailed proof, but refer the reader to our Isabelle implementation.


\subsection{Equivalence}

To show that the model-theoretic and fixed-point semantics of \core{}
compute the same Herbrand model, we need to show that the least fixed
point of the immediate consequence operator is equal to the
intersection of all the Herbrand models for any program.

\begin{lemmah}
Herbrand models of a \core{} program are fixed points of the immediate consequence operator.    
\end{lemmah}

\begin{lemmah}
Fixed points of the immediate consequence operator of a \core{} program that are subfact-closed are Herbrand models of the program.    
\end{lemmah}

By proving that the Herbrand models and subfact-closed fixed points of
the immediate consequence operator are the same, we conclude that the
least fixed point of the immediate consequence operator
$\textit{IC}_P$ (a subfact-closed database) is equal to the
intersection of all its Herbrand models.

\begin{theoremh}
\label{theorem:main}
The model theoretic semantics and fixed point semantics of \core{} are equivalent.
\end{theoremh}


\section{Applications of First-Class Facts}
\label{sec:apps}

\revised{We now ground our exploration of \dls{} in some familiar frameworks. To begin with, we observe that \dls{} strictly extends Datalog: any Datalog program is a terminating \dls{} program (and vice-versa, though with a potential complexity blowup). However, the true power of \dls{} lies in its ability to express queries wherein first-class facts are recursively (a) used as triggers and (b) generated by the computation. In effect, the combination of first-class rules and recursion enables ad-hoc polymorphic rules, which enable an expressive programming style similar to functional programming and natural deduction. We explore this power by a series of applications including provenance, algebraic data types, functional programming, structural abstract interpretation, and type systems.}


\subsection{Datalog in \dls{} and \ldls{}}
\label{sec:apps-dl}
\revised{\begin{definition}[{Datalog programs in \dls{} and \ldls{}}]
\label{def:datalog}
The Datalog subset of \dls{} is a subset wherein all occurrences of the binders for \textit{id}s in the body are wildcards. 
\end{definition}
\begin{theorem}[Datalog Programs Terminate]
If $P$ is a Datalog program in \dls{}, or \ldls{}, then the program terminates.
\begin{proof}
For \core{}, termination follows directly from the fact that the restricted chase for Datalog programs without existentials necessarily has a finite number of iterations: because the bodies of rules $\phi$ in Algorithm~\ref{alg:chase} may not bind \textit{id} columns (only wildcards), there is no ability to build a Skolem term of non-trivial height. Thus, as in Datalog, the restricted chase for \core{} terminates. The termination for \ldls{} follows similarly: because the body may not bind any \textit{id}s in the variables $x_i$, the height of terms produce by \textit{subfact} may not increase throughout iterated application of $\textit{IC}_R$. 
\end{proof}
\end{theorem}}



\subsection{Provenance in \dls{}}
\label{sec:why-where}

\revised{Database provenance refers to the concept of tracking the origin of a database record.
In Datalog and knowledge-graph reasoning, the coarsest and most popular form of provenance is lineage (why-provenance) which, for each tuple in the output (IDB), identifies a set of contributing tuples from the input (EDB)~\cite{Cui:2000,cheney2009provenance}.
Typical implementations of lineage eagerly annotate each tuple with a tag during query computation to track its origin, and Datalog rules are rewritten to propagate the derivation order of these tags. After the query finishes, lineage is collected by computing the downward closure on the derivation graph, gathering all reachable leaf nodes to explain the tuple’s origin. Such annotations are naturally isomorphic to the \textit{id} values generated when evaluating existentially quantified queries in \dls{}.
Meanwhile, \dls{}’s ability to directly construct nested facts naturally facilitates eager derivation computation. The derivation graph can be represented as a binary relation, \textit{deriv}, which tracks all tuple \textit{id}s from the body clauses of a Datalog rule to its head clause.
For example, the derivation of the rule:
\[
H(a, c) \leftarrow B_{0}(a, b) \wedge B_{1}(b, c)
\]
can be a computed via:
\[
\begin{array}{rl}
    \exists! \textit{id}_{3}, \textit{id}_{4}. & \!\!\!\!  \textit{id}_{3} = \textit{deriv}(\textit{id}_0, \textit{id}) \wedge \textit{id}_{4} = \textit{deriv}(\textit{id}_1, \textit{id} ) \leftarrow  \\
      & \!\!\!\! \!\!\!\! \textit{id}_0 = B_{0}(a, b) \wedge \textit{id}_1 = B_{1}(b, c) \wedge \textit{id} = H(a, c).
\end{array}
\]
After constructing the derivation graph, the downward closure can be efficiently computed as a reachability query from the IDB tuple \textit{id} to the EDB, identifying all contributing input tuples.}

\revised{The eager approach stores complete provenance information for every tuple in the Datalog relation. Once the rule is computed, further analysis can be easily performed by querying the generated provenance relation. However, eagerly tracking provenance in datalog is data-intensive, with its time complexity proven to be NP~\cite{10.1145/3651146}. In some specific applications, such as debugging, only the provenance of a few specific tuples is required.
A cheaper but incomplete alternative is to compute the provenance in an lazy approach (also called on-demand approach). Instead of generating the full provenance during query evaluation, this method computes provenance only for the specific tuple that needs explanation or debugging. This involves extracting constraint formulas from the input and output databases and solving them using techniques such as SAT~\cite{calautti2024computing} or semi-ring solvers~\cite{esparza2015fpsolve}.}

\revised{Although \dls{}'s semantics eagerly annotates all tuples---na\"ively yielding eagerly-materialized provenance---a hybrid approach can still be adopted by using a first-class fact to trigger lineage computation, similar to lineage tracking in the WHIPS system~\cite{cui2000lineage}. To lazily materialize the lineage, we may modify the previous rule: 
\[
\begin{array}{rl}
     \exists! \textit{id}_{5}, \textit{id}_{6}. & \!\!\!\! \textit{id}_{5} = \textit{explain\_t}(\textit{id}_0) \wedge \textit{id}_{6} = \textit{explain\_t}(\textit{id}_1) \leftarrow  \\
      & \!\!\!\! \!\!\!\! \textit{id}_{0} = B_{0}(a, b) \wedge \textit{id}_1 = B_{1}(b, c) \wedge \textit{id}_{3} = H(a, c) \wedge \\
      & \!\!\!\!  \!\!\!\! \textit{id}_{4} = \textit{explain\_t}(id).
\end{array}
\]
Here, the \textit{explain\_t} relation constrains the search space. It initially stores the single tuple that needs to be explained, and during the computation, it is gradually expanded to include all tuples that cover the nodes in the derivation paths leading to the initial tuple. 
After \textit{explain_t} has been propagated, the EDB relation IDs stored within it will be the why-provenance associated with the initial tuple needing explanation.}

\revised{While why-provenance tracks the origin of each output tuple by annotating entire rows, where-provenance goes further by annotating individual columns~\cite{buneman2001and}. This approach identifies the exact source locations (i.e., specific columns in the input data) that contribute to each value in the output, allowing for fine-grained tracing of data lineage at the attribute level.}


\revised{In \dls{}, a na\"ive way to capture where-provenance is to associated each concrete column value in a EDB relation with a tuple in \textit{column} relation, whose id is isomorphic to the column annotation in definition of where-provenance.
For example, the where-provenance $\textit{prov}_H$ of relation $H$ in query $H(a, c) \leftarrow B_{0}(a, b) \wedge B_{1}(b, c)$ can be materialized as:
\[
\begin{array}{rl}
  \exists!  & \!\!\!\! \!\! \textit{id}_H. \textit{id}_H = \textit{prov}_H(\textit{id}_{a0}, \textit{id}_{c1})  \leftarrow  \\
    & \textit{id}_0 = B_0(a, b) \wedge \textit{id}_1 = B_1(b, c) \\
    & \textit{id}_{a0} = \textit{column}(\textit{id}_0, 0, a) \wedge \textit{id}_{b0} = \textit{column}(\textit{id}_0, 1, b) \wedge \\
    & \textit{id}_{b1} = \textit{column}(\textit{id}_1, 0, b) \wedge \textit{id}_{c1} = \textit{column}(\textit{id}_1, 1, c)
\end{array}
\]
Here, \textit{column} is a ternary relation, where $\textit{column}(i, k, v)$ indicate  the $k$-th column value of tuple associate with identity $i$ is $v$. } 





\subsection{Algebraic data types}
\label{sec:adts}

Some Datalogs have a dedicated system for algebraic data types (ADTs) supporting records, unions, and tagged variants. For example, Souffl\'e offers a useful ADT system with heap-allocated structured values. In Souffl\'e, a heap-allocated structure is denoted with a \$ sign; so, the following example is a Souffl\'e rule that \revised{says} if a lambda exists as a subexpression of any parent expression ($\_$), then its body \texttt{eb} is a subexpression of the lambda abstraction:

{\small
\begin{Verbatim}[baselinestretch=0.75,commandchars=\\\{\}]
 \rtag{subexpr}(\$lam(x, eb), eb) :- \rtag{subexpr}(_, \$lam(x, eb)).
\end{Verbatim}
}

Unfortunately, ADTs cannot trigger rule evaluation and are not indexed---which has a performance impact we explain further in \S\ref{sec:eval}. For this reason, Souffl\'e requires we assert that the lambda is a subexpression of any expression since the \revised{lambda-syntax} value must be referenced from some fact before it can be manipulated by a rule. Souffl\'e implements this rule as a scan of the \rtag{subexpr} relation, extraction of its second-column value, a lookup of that value's second field, a guard to check that it is a lambda, and finally, insertion of the lambda paired with its body-reference \texttt{eb} in the \rtag{subexpr} relation. 
By contrast, typical Datalog evaluation relies on relational joins, which do not normally require any guard but are implemented using an efficient B-tree lookup to determine the exact relevant tuples for the join. Following is an example that builds a transitive \rtag{within} relation from this \rtag{subexpr} relation. It is implemented in Souffl\'e using a relational union, a scan and insert, for the first rule, and a binary join (fused with a projection of the shared column) for the second rule. Operationally, this join would scan the outer relation \rtag{within}, lookup only those \texttt{ec} values that are paired with the relevant \texttt{ei} in \rtag{subexpr} using its tree-based index, and then insert each unique \texttt{e,ec} tuple into \rtag{within}.
Here (Souffl\'e code), \rtag{within} is computed as the transitive closure of \rtag{subexpr}:

{\small
\begin{Verbatim}[baselinestretch=0.75,commandchars=\\\{\}]
   \rtag{within}(e, ec) :- \rtag{subexpr}(e, ec).
   \rtag{within}(e, ec) :- \rtag{within}(e, ei), \rtag{subexpr}(ei, ec).
\end{Verbatim}
}
In Souffl\'e, since ADTs are heap allocated and not indexed the way (top-level) facts are, these values cannot be efficiently selected for in the course of a relational join operation. Because Souffl\'e cannot easily index its ADTs, a scan-and-filter approach is used.
Consider a rule that computes a relation \rtag{shadows} pairing each lambda with each other lambda that defines a shadowing variable. 

{\small
\begin{Verbatim}[baselinestretch=0.75,commandchars=\\\{\}]
   \rtag{shadows}(\$lam(x, eb0), \$lam(x, eb1)) :- 
        \rtag{subexpr}(\_, \$lam(x, eb0)), \rtag{within}(eb0, e), 
        \rtag{subexpr}(e, \$lam(x, eb1)).
\end{Verbatim} 
}

Operationally, this is implemented as a three-way join between \rtag{subexpr}, \rtag{within}, and \rtag{subexpr} (which takes advantage of indexes for these relations), introspecting on the outer lambda to determine its body value (\texttt{eb0}) during this join, followed by introspection on the inner lambda to determine whether both \texttt{x} values match, before inserting relevant pairs of lambdas into \rtag{shadows}. This implicitly materializes the Cartesian product of nested lambda values encountered during the join (in time, not space) before filtering to only insert those with matching formal parameters. Because \texttt{\$lam} values are scattered in memory and only referenced from top-level relations and are not collectively indexed, there is no straightforward way to efficiently select just those \texttt{\$lam} values that are relevant to a particular \texttt{x} as we could when selecting \rtag{subexpr} values with a particular \texttt{e}. The code generated by Souffl\'e for a similar case, from a program analysis, is shown in Figure~\ref{fig:souffle-cpp}.

\subsection{Functional Programming}
\label{sec:fp}

\revised{With first-class facts that \emph{can trigger} rule evaluation, and that are indexed automatically and amenable to efficient joins, natural idioms emerge for performing efficient functional programming within Datalog, that will benefit from data-parallel deduction as detailed in Section~\ref{sec:slog}. First-class facts permit Reynolds' \emph{defunctionalization}~\cite{reynolds1972definitional}, a classic transformation from higher-order functions to first-order functions with ad hoc polymorphism over structured values. Note that first-class facts gives us precisely the features we need to implement this: structured data that can trigger rule evaluation via (naturally polymorphic) joins on their identity!}

\revised{To illustrate this, consider the $\lambda$-calculus interpreter presented in Figure~\ref{fig:lambdacalc}. The upper left shows the traditional big-step (natural-deduction style) rules of inference, that define relation $(\Downarrow)$, for evaluating variable references, lambda abstractions, and lambda applications. The right side shows a corresponding interpreter in \ldls{}. The bottom left gives rules for defunctionalized higher-order environments built using a chain of $(\mapsto)$ facts. E.g., the first-class map $[ 0 \mapsto 1, 2 \mapsto 3]$ can be constructed as $\mapsto\!\!(\mapsto\!\!(\bot(), 0, 1), 2, 3)$, using these rules, and then accessed efficiently via \textit{lookup}.}

\revised{A 3-way join between this \textit{lookup} relation, a syntax-encoding \textit{ref} relation, and an \textit{eval} relation, implements variable reference in our interpreter. The \textit{eval} relation encodes an input $e,\rho$ pair so the interpreter can be demand driven; the $(\Downarrow)$ relation associates an \textit{eval} input with its output, the value it denotes. The second rule joins the \textit{eval} and $(\lambda)$ relations, and builds closures, \textit{clo} facts, that are associated with the \textit{eval} fact in $(\Downarrow)$. The final \textsc{App} rule is encoded as four \ldls{} rules because it makes three recursive uses of the interpreter. Our implementation of \ldls{} as a full language, \slog{} (discussed in Section~\ref{sec:slog}) provides syntactic sugar that will perform this transformation of a single rule into multiple rules for the user automatically, but in this paper we present it desugared to focus on the key principles at play. The rule shown with a conjunction in the head is really two \ldls{} rules that infer both subexpressions of an application must be evaluated before the application itself. The bottom rule deduces that once these subexpressions have been evaluated, the body of the function being applied must be evaluated under an extended environment that binds the formal parameter ($x$) to the argument value ($v_a$) using a $(\mapsto)$ fact. Finally, the top rule (shown parallel to \textsc{App} on the left) puts this all together and deduces that the application expression's value is the same as that of the function body under the appropriate environment.}

\revised{IncA, the other recent approach to encoding functional programming in Datalog~\cite{pacak2022functional, pacak:20} requires static monomorphization as a compilation approach because it is not able to directly leverage first-class facts to perform proper defunctionalization. Our extension, \dls{}, grants us an efficient implementation approach that may be extended to a scalable system for performing data-parallel functional programming in Datalog. Sections~\ref{sec:slog} and \ref{sec:eval} present further details of our practical system and evaluate it on applications that leverage the functional techniques we've discussed.}

\begin{figure}
\begin{flushleft}
\fbox{$e, \rho \Downarrow v$}
\end{flushleft}
\vspace{-0.3cm}

\begin{tabular}{c|c}
{{\small\textsc{Var}}\quad{\LARGE ${ \frac{v\,=\,\rho(x)}{x, \rho\ \Downarrow\ v}}$}} \quad&
\begin{minipage}{2.5in}
\begin{Verbatim}[baselinestretch=.8,commandchars=\\\{\},codes={\catcode`$=3\catcode`^=7}]
$\Downarrow\!(\textit{call}, v) \leftarrow$
  $\textit{call} = \textit{eval}(\textit{ref}(x),\rho),$
  $\_\!=\!\textit{lookup}(\rho, x, v)$
\end{Verbatim}
\end{minipage}
\\
\\
{{\small\textsc{Abs}}\quad{\LARGE ${ \frac{ }{(\lambda\,(x)\,e),\rho\ \Downarrow\ \langle(\lambda\,(x)\,e), \rho\rangle}}$}} \quad& 
\begin{minipage}{2.5in}
\begin{Verbatim}[baselinestretch=.8,commandchars=\\\{\},codes={\catcode`$=3\catcode`^=7}]
$\Downarrow\!(\textit{call}, \textit{clo}(\textit{lam}, \rho)) \leftarrow$
  $\textit{call} = \textit{eval}(\textit{lam},\rho),$
  $\textit{lam} = \lambda(x, e)$
\end{Verbatim}
\end{minipage}
\\
\\
{{\small\textsc{App}}\quad{\LARGE ${ \frac{\overset{\mathlarger{e_f,\rho \Downarrow \langle(\lambda\,(x)\,e_b), \rho_\lambda\rangle}}{e_a,\rho \Downarrow v_a \quad e_b,\rho_\lambda[x \mapsto v_a] \Downarrow v}}{ (e_f~e_a),\rho\ \Downarrow\ v }}$ }}\quad& 
\begin{minipage}{2.5in}
\begin{Verbatim}[baselinestretch=.8,commandchars=\\\{\},codes={\catcode`$=3\catcode`^=7}]
$\Downarrow\!(\textit{call}, v) \leftarrow$
  $\textit{call} = \textit{eval}(\textit{app}(e_f\,e_a), \rho),$
  $\_\!=\,\Downarrow\!(\textit{eval}(e_f,\rho),$
       $\textit{clo}(\lambda(x,e_b), \rho_\lambda)),$
  $\_\!=\,\Downarrow\!(\textit{eval}(e_a,\rho), v_a),$
  $\_ =\,\Downarrow\!(\textit{eval}(e_b,\rho', v),$
  $\rho' =\ \mapsto\!\!(\rho_\lambda, x, v_a))$
\end{Verbatim}
\end{minipage} 
\\
\\
\begin{minipage}{1.6in}
{\small\textsc{Defunctionalized Environments}}
\begin{Verbatim}[baselinestretch=.8,commandchars=\\\{\},codes={\catcode`$=3\catcode`^=7}]

      $\textit{lookup}(\rho, x, v) \leftarrow\ $
        $\rho =\ \mapsto\!\!(\_, x, v)$
  
      $\textit{lookup}(\rho, x, v) \leftarrow $
        $\rho =\ \mapsto\!\!(\rho', y, \_),$
        $x \neq y,$
        $\textit{lookup}(\rho', x, v)$
\end{Verbatim}
\end{minipage} 
& 
\begin{minipage}{2.5in}
\begin{Verbatim}[baselinestretch=.8,commandchars=\\\{\},codes={\catcode`$=3\catcode`^=7}]
$\textit{eval}(e_f,\rho), \textit{eval}(e_a,\rho) \leftarrow$
  $\_\!=\!\textit{eval}(\textit{app}(e_f\,e_a), \rho)$

$\textit{eval}(e_b,\mapsto\!\!(\rho_\lambda, x, v_a)) \leftarrow$
  $\_\!=\!\textit{eval}(\textit{app}(e_f\,e_a), \rho),$
  $\_\!=\,\Downarrow\!(\textit{eval}(e_f,\rho),$
       $\textit{clo}(\lambda(x,e_b), \rho_\lambda)),$
  $\_\!=\,\Downarrow\!(\textit{eval}(e_a,\rho), v_a)$
\end{Verbatim}
\end{minipage} 
\\
\end{tabular}

\caption{\revised{Lambda-calculus interpreter $(\Downarrow)$ in \ldls{}.}}
\label{fig:lambdacalc}
\end{figure}

\subsection{Structural Abstract Interpretation}
\label{sec:aam}

A principal application motivating recent developments in Datalog has been high-performance declarative simulations of complex systems, especially software. Static program analyses aim to model software soundly based on the program text alone, with sufficient precision to prove meaningful correctness, security, observational equivalence, information flow, and other properties. 
Over the past decade or so, a line of work has gone into systematic and tunable approaches to applying \emph{abstract interpretation} (AI)~\cite{cousot77unifiedmodel,cousot1996abstract,cousot1979systematic} to structural operational semantics~\cite{hennessy1990semantics, plotkin1981structural} and natural semantics~\cite{kahn1987natural} via a structural abstraction that is as straightforward as possible. 
The \emph{abstracting abstract machines} (AAM) methodology~\cite{might2010abstract,VanHorn:2010} prescribes a particular systematic application of abstract interpretation on abstract-machine operational semantics like those we built in \S\ref{sec:fp}---this approach is a natural one for traditional Datalog-based analyses as it requires a flat, small-step semantics with a particular abstraction of the stack (once explicitly modeled in the semantics). Some AAM-based approaches have encapsulated the abstraction in a tunable monadic interpreter~\cite{sergey2013monadic}.
A related trend toward \emph{abstracting definitional interpreters} (ADI)~\cite{darais2017abstracting, brandl2023modular} follows a similar approach, except applies AI directly to naturally recursive definitional interpreters, sometimes using a monad transformer stack for tunable abstraction and a partial-evaluation-based approach to achieve performance in the 
implementation \cite{amin2017collapsing, wei2018refunctionalization, wei2019staged}.

The literature on the AAM approach shows that a \emph{store-passing} transformation (as might also be used to model, \emph{e.g.},  direct mutation) can handle indirect recursion in the \emph{abstract} domains (\emph{e.g.}, binding environments contain closures which contain environments) through a set of references that are finitized under AI~\cite{might2010abstract}. This forms a key preparatory transformation enabling a straightforward homomorphic structural abstraction to follow.
For any abstract-machine component referenced through the store, the \emph{allocation} of its \emph{abstract addresses} becomes an expressive proxy for tuning the polyvariance (e.g., context sensitivity, flow sensitivity) of those analysis components, and one with the notable property that soundness can be guaranteed for all tunings of abstract allocation~\cite{gilray2016poly, might2009posteriori}.
This is likewise the natural approach for continuation-passing abstract machines that store-allocate continuations at function calls~\cite{VanHorn:2010}, and is a key enabling technique as AAMs require some way of finitizing otherwise-unbounded recursive evaluation. A particular reflective choice of address for continuations (the function entry point, in the analysis, including its abstract environment, after parameters are bound) can guarantee the analysis is equivalent to a pushdown system with an unbounded stack~\cite{gilray2016p4f}.
A wide variety of heavyweight verification tasks can be built atop this foundation, 
including higher-order pointer and shape analysis~\cite{might2008exploiting,germane2020liberate, germane2021newly}
and abstract symbolic execution~\cite{tobin2012higher,nguyen2017soft}.


\input{figure-mcfa}

\revised{We implement a core AAM-based analysis that permits an apples-to-apples comparison between our system \slog{} that implements \dls{} (see \S\ref{sec:slog}), and Souffl\'e (see \S\ref{sec:adts}). It is derived from a stack-passing interpreter---a call-by-value Krivine's machine~\cite{krivine:2007:cbn}---with a store factored out. Figure~\ref{fig:kcfa-mcfa} shows our small-step control-flow analysis (CFA) in Souffl\'e with a global store and a tunable instrumentation.
The original $k$-CFA \cite{shivers:1991:diss} used true higher-order environments, unlike equivalent analyses written for object-oriented languages which used flat objects. Our analysis in Figure~\ref{fig:kcfa-mcfa} is implemented as the corresponding CFA for functional languages, called $m$-CFA~\cite{might2010resolving}, 
and is used in our evaluation to compare first-class fact vs ADT handling in \slog{} vs Souffl\'e (see \S\ref{sec:eval:aam}).}

\begin{figure}[t!]
\begin{flushleft}
\fbox{$\Gamma \vdash e : \tau$}
\end{flushleft}
\begin{tabular}{c|c}
 {{\small\textsc{T-Var}}\quad{\LARGE ${ \frac{x : T \, \in\, \Gamma}{\Gamma\, \vdash\, x : T}}$}} \quad&
\begin{minipage}{2.5in}

\begin{Verbatim}[baselinestretch=.8,commandchars=\\\{\},codes={\catcode`$=3\catcode`^=7}]
$\textit{id}_0 = \textit{ck}(\textit{ref}(x), \Gamma),$
$\_ = \textit{lookup}(\Gamma, x, T)$
$\rightarrow \textit{type}(\textit{id}_0, T)$
\end{Verbatim}
\end{minipage}
\\
\\ 
{{\small\textsc{T-Abs}}\quad{\LARGE ${ \frac{\Gamma, x\, : \,T_1 \, \vdash \, e \, : \, T_2}{(\lambda\,(x\,:\, T_1)\,e)\, : T_1 \rightarrow \,T_2}}$}} \quad& 
\begin{minipage}{2.5in}
\begin{Verbatim}[baselinestretch=.8,commandchars=\\\{\},codes={\catcode`$=3\catcode`^=7}]
$\textit{id}_0 = \textit{ck}(\lambda (x, T_1, e),\Gamma),$
$\_ \ \ \ = \textit{type}(\textit{ck}(e, \rho'), T_2)$
$\rho' \ =\ \mapsto\!\!(\Gamma,x,T_1)$
$\rightarrow \textit{type}(id_0, \textit{arrow}(T_1, T_2))$

\end{Verbatim}
\end{minipage}
\\
 
{{\small\textsc{T-App}}\quad{\LARGE ${ \frac{{\Gamma \, \vdash \, e_0\,:\,T_0\, \rightarrow\, T_1}\quad e_1\,:\,T_0}{\Gamma \, \vdash \, (e_0~e_1)\,:\,T_1}}$ }}\quad& 
\begin{minipage}{2.5in}
\begin{Verbatim}[baselinestretch=.8,commandchars=\\\{\},codes={\catcode`$=3\catcode`^=7}]
$\textit{id}_0 = \textit{ck}(\textit{app}(e_0, e_1), \Gamma),$
$\_ = \textit{type}(\textit{ck}(e_0, \Gamma), \textit{arrow}(T_0,T_1)),$
$\_ = \textit{type}(\textit{ck}(e_1, \Gamma), T_0),$
$\rightarrow \textit{type}(\textit{id}_0, T_1)$
\end{Verbatim}
\end{minipage} \\
\end{tabular}

\caption{\revised{STLC type rules (left); equivalent \ldls{} (right).}}
\label{fig:stlc}
\end{figure}
\subsection{Type Systems}
\label{sec:types}

\revised{Structural type theories provide a formal framework for ensuring that programs are free from runtime type errors~\cite{tapl}. Such systems grew out of a rich history of constructive logic (e.g., Per Martin-L\"of's intuitionistic type theory~\cite{Martin-Lof1996}), typically specifying typing rules via natural deduction, and form the basis of type checking and inference systems in myriad modern programming languages, such as OCaml and Haskell. Implementing type checkers for structural type systems involves iteratively traversing a program's abstract-syntax tree and attempt to assemble a valid type derivation which adheres to the typing judgement for the given language.}

\revised{Traditionally, structural type systems have been implemented in functional programming languages, using recursion and pattern matching. However, Datalog has attracted recent attention in implementing type systems due to its logic-defined nature, along with the potential for automatic incrementalization and high-performance compilation~\cite{Pacak:2019, pacak:20}. These efforts are challenged by Datalog's semantic restrictions: type systems often involve traversing tree-shaped abstract syntax trees and assembling tree-shaped typing derivations. As such, Datalog-based type systems require tedious refactoring of the type system, carefully compiling type checking algorithms into rules which respect Datalog's restrictions via techniques such as monomorphization, ultimately yielding implementations which are far removed from the original typing rules.}

\revised{We have found that \dls{} has proven a promising implementation candidate for structural type systems, and have used \ldls{} to implement numerous type checkers for a range of systems including first-order dependent types~\cite{atapl} and intuitionistic type theory~\cite{Martin-Lof1996}. As an example, Figure~\ref{fig:stlc} shows a transliteration of the Simply-Typed Lambda Calculus (STLC) into \ldls{}, following a textbook presentation~\cite{tapl}. The left-hand side of the figure displays the three type rules for STLC. The first says that if $x$ has type $T$ in typing environment $\Gamma$, then $x$ can be checked to the type $T$ in $\Gamma$. This is mirrored by the \ldls{} rule on the right, which uses the \textit{ck} fact to trigger lookup of $x$ (using the defunctionalized environments from the bottom of Figure~\ref{fig:lambdacalc}) and materialization of the \textit{type} judgement. The second rule triggers when a lambda is checked, subsequently extending the environment and checking the type of the body, finally assembling an function (\textit{arrow}) type. Last, \textsc{T-App} checks an application, ensuring that the function's input ($T_0$) matches with $e_1$'s type. In STLC, the simple nature of types means that \dls{}'s structural equality suffices for type comparison' More complex type systems equate types, would also implementing a function to (\emph{e}.\emph{g}.,) canonicalize types. Last, we omit rules to trigger \textsf{ck} for subordinate expressions for brevity; our implementation uses syntactic sugar to enable writing a single rule for each of these cases. }

\section{\slog{}: Implementing \ldls{}}
\label{sec:slog}

\revised{In this section, we describe the implementation of \ldls{} as a practical language.
\emph{Structured Datalog} (\slog{}) represents a language design and implementation methodology based on \dls{},} Datalog extended with the principle that all facts are also first-class structured data and every structured datum is a first-class fact.
We have implemented \slog{} in a combination of Racket (the compiler,
roughly 10,600 new lines), \cpp (the runtime system and parallel RA
backend; roughly 8,500 new lines), Python (a REPL and daemon; roughly
2,500 new lines), and in \slog{} (60 lines for language feature support). 
In \slog{}, fact identities are implemented as 64-bit primary keys implemented via a distributed and parallel \textsf{autoinc}.

Our Racket-based compiler translates \slog{} source code to \cpp{} 
that links against our parallel relational-algebra backend. Our compiler is 
structured in the nanopass style~\cite{nanopass}, composed of a larger number 
of small passes which consume and produce various forms of increasingly-low-level 
intermediate representations (IRs). After parsing, an organize 
pass performs various simplifications, such as canonicalizing the direction of 
implication (\texttt{-{}-{}>}) and splitting disjunctive body clauses and 
conjunctive head clauses into multiple rules. This pass also eliminates various 
syntactic niceties of the language.
Subsequent passes perform join planning, then index generation: computing a 
minimal necessary set of  shared indices, ensuring there is always an 
interning index that models the relation as a mapping from
fact-tuple values to an \emph{unsigned 64-bit integer encoding a unique 
intern id for that fact}. The compiler then performs Strongly Connected Component (SCC) computation with 
Tarjan's algorithm~\cite{tarjan-scc} and stratifies grounded fact-negation 
operations and aggregation operations. Then, the compiler performs 
incrementalization, or the so-called semi-na\"ive transform in which rules
are triggered by the delta of a relation accessed as a body clause---reducing
the substantial recomputation of so-called na\"ive fixpoint evaluation. 
Finally, the compiler writes out a set of optimized and pass-fused 
relational algebra operations (e.g., join followed by projection) in \cpp.

\begin{figure}[t]
\centering
\includegraphics[width=\linewidth]{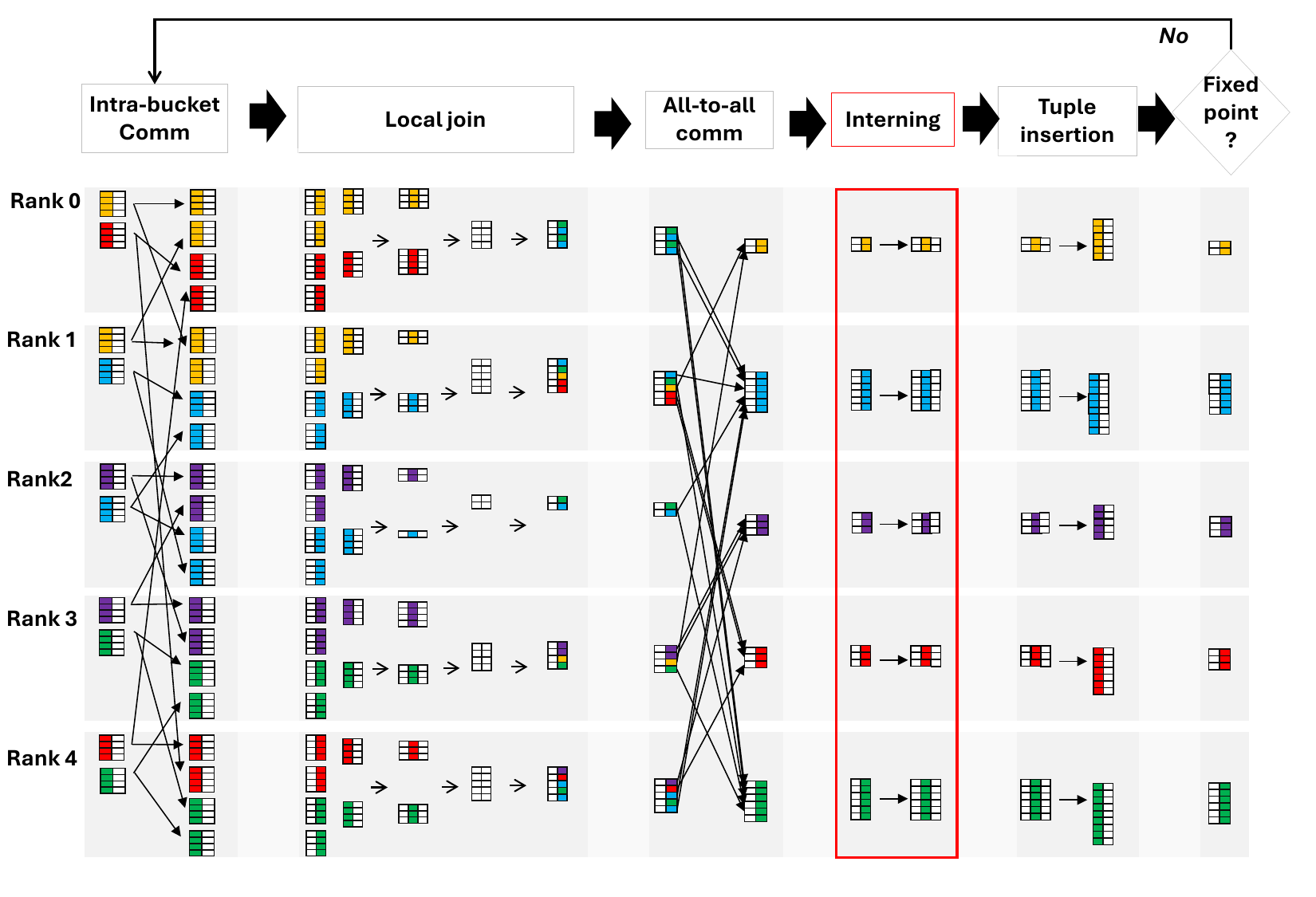}
\caption{An illustration of the main phases of our parallel RA backend.}
\label{fig:sid}
\end{figure}

\subsection{MPI-based Data-parallel Implementation}
\label{sec:mpi}

We extend an MPI-based Datalog back-end from our prior work on data-parallel 
relational algebra~\cite{cc,kumar:hipc:2019,kumar:2020}, to demonstrate that the key extension of Datalog we propose
is naturally data-parallel, fact identity being derived from reads over indices
in precisely the same data-parallel manner as rule evaluation in Datalog.

Our parallel relational-algebra (RA) backend supports fixed-point
iterations and is designed for large-scale multi-node HPC
clusters. Based on the bulk-synchronous-processing protocol and built
using the \texttt{MPI-everywhere} model~\cite{forum1994mpi, zambre2021logically},
the parallel RA framework addresses the problem of partitioning and balancing workloads across
processes by using a two-layered distributed
hash table~\cite{kumar:hipc:2019}. In order to materialize newly generated facts within each iteration, and thus facilitate \emph{iterated} RA (in a fixed-point loop), an all-to-all data exchange phase is used at every iteration. 

Figure~\ref{fig:sid} shows a schematic diagram of all the phases of our parallel relational algebra, including a new \emph{interning} phase, in the context of \slog{}'s fixed-point loop. \revised{ There are five primary phases of our system: intra-bucket communication, local RA computation, all-to-all data exchange, interning, and materialization in appropriate indices. Relations are partitioned across processes using a modified double-hashing approach~\cite{Valduriez:1988:PET:54616.54618}.
This involves partitioning relations by a hash of their join-column values so that they can be efficiently distributed to participating processes. The main insight behind this approach is that for each tuple in the outer relation, all relevant tuples in the inner relation must be hashed to the same bucket, stored on the same MPI process or node, permitting joins to be performed locally on each process. To handle any possible key-skew (imbalance across keys) in the relations themselves, we also hash the non-join columns to map the tuple to a sub-bucket. Each unique bucket/sub-bucket pair is then mapped to an MPI process. To distribute subbuckets to managing processes, we use round-robin mapping and have a mechanism to dynamically refine buckets that accumulate large numbers of tuples into a greater number of sub-buckets.}

\paragraph{Intra-bucket Comm. and Local Join} 
\revised{Our double-hashing approach to manage key-skew necessitates an intra-bucket communication phase to co-locate matching tuples before the join: all sub-buckets for the outer relation, $R_\Delta$, are transmitted to all other sub-buckets (in the same bucket) temporarily for the join. 
During the local computation phase, RA kernels (comprised of fused join, selection, projection, and union) are executed in parallel across all processes.}

\paragraph{All-to-all Comm.}
\revised{Output tuples generated from local joins may each belong to an arbitrary bucket in the output relation, so a non-uniform all-to-all communication phase shuffles the output to their managing processes (preparing them for any subsequent iterations).
Materializing a tuple in an output relation involves hashing on its join and non-join columns to find its bucket and sub-bucket (respectively), and then transmitting it to the process that maintains that bucket/sub-bucket.
The overall scalability of the RA backend relies on the scalability of the all-to-all inter-process data exchange phase, but all-to-all is notoriously difficult to scale~\cite{4536141, scott1991efficient, thakur2005optimization}---largely because of the \emph{quadratic} nature of its workload. 
We address this scaling issue by adopting recent advancements~\cite{fan:2022} that optimize non-uniform all-to-all communication by extending the $\log$-time Bruck algorithm~\cite{bruck1997efficient, thakur2005optimization, traff2014implementing} for non-uniform all-to-all workloads.}

\revised{\paragraph{Deduplication, Interning, and Insertion}
Once new tuples are received, we perform the interning phase required to assign unique first-class facts their unique identity. This first checks if the received fact was already discovered; if not, then a new 64-bit intern id is created and associated with the fact. This is done by reserving the first 16 bits of the 64-bit ID for the relation ID, the next 16 bits for the bucket ID, and the remaining 32 bits for a unique fact ID---generated on a per-process basis via simple bump-pointer allocation. Critical to our approach, when facts are generated, they are sent to the canonical index, so that fact-ID generation may happen in parallel on each bucket, as each bucket performs local insertion and deduplication into the canonical (master) index, without any additional communication or synchronization required.
Each fact is then inserted into $R_\text{new}$, and following the semi-na\"ive evaluation approach, these newly generated facts form the input $R_\Delta$ for the following iteration of the fixed-point loop. This process continues until a fixed point is reached.}

\section{Applications and Evaluation}
\label{sec:eval}
In this section, we elaborate upon several applications of \slog{}; we directly evaluate the performance of \slog{} against several state-of-the-art tools in the context of provenance and program analysis. 


\subsection{Eager Why-Provenance (Lineage)}
\label{sec:eval-why}


In prior literature, why-provenance has been deemed useful but eagerly computing it is considered too expensive because it requires storing all possible derivations of a tuple. By contrast, the on-demand approach calculates possible derivations for user-specified tuples after the execution. For this reason, existing Datalog engines compute why-provenance on-demand in a top-down fashion, thwarting effective data parallelism. \slog{}'s parallel capabilities prompt us to again investigate eager provenance evaluation.

We explore this by comparing the running time for eager lineage calculation using \slog{} and three other Datalog engines: VLog, Nemo, \revised{and RDFox}~\cite{urbani2016column, ivliev2023nemo, motik2014parallel}. For VLog, we use its Java version, Rulewerk, which supports lineage computation via an existential operator. Nemo is another engine, similar to VLog, but implemented in Rust. RDFox is a high-performance commercial Datalog engine that supports lineage computation via a built-in Skolem function. These experiments were done on a machine with a 3.0Ghz 12-core AMD 5945WX (Zen3 Chagall) with 128GB of memory. We ran each experiment five times and report the mean; results are shown in Table~\ref{table:whyprovenance}. We provide single-thread performance comparisons for all engines, and for \revised{parallel RDFox and} \slog{}, we also report the running times when all cores on our test platform are utilized.

\begin{table}[h]
\caption{Running Time (s): Eager why-provenance vs. Nemo, Rulewerk (single threaded), RDFox; timeout (\timeout{}) of one hour.}
\label{table:whyprovenance}
\resizebox{\linewidth}{!}{
\begin{tabular}{@{}c|c|cccccccc@{}}
\toprule
\textbf{Query} & \textbf{Dataset} & \textbf{Nemo} & \textbf{Vlog} &  \multicolumn{2}{c}{\revised{\textbf{RDFox (threads)}}}  & \multicolumn{2}{c}{\textbf{\slog{} (threads)}} \\ 
               &                  & (Rust)         & (Java)       &  \revised{1}     &  \revised{12}     & 1  & 12  \\ \midrule
\multirow{3}{*}{\rotatebox{90}{Galen}} 
               & 15               & 1.63          & 7.38          &   \revised{0.20}     &  \revised{0.14}   & 0.68       & \textbf{0.13}  \\
               & 25               & 2.35          & 17.7          &   \revised{0.33}     &  \revised{0.19}   & 0.92       & \textbf{0.17}  \\
               & 50               & 34.5          & 415           &   \revised{2.30}     & \revised{\textbf{1.98}}   & 12.0            & 2.19  \\ 
               \midrule
\multirow{3}{*}{\rotatebox{90}{CSDA}}  
               & httpd            & 72.2          & 152           &  \revised{57.9}    &  \revised{40.7}  & 127         & \textbf{22}    \\
               & linux            & 548           & \timeout{}    &  \revised{315}     &  \revised{230} & 6,237       & \textbf{116}  \\
               & postgres         & 234           & \timeout{}    &  \revised{138}     &  \revised{95.2}  & 320    & \textbf{62.6}  \\ 
               \midrule
\multirow{4}{*}{\rotatebox{90}{Andersen}} 
               & 10,000            & 1.38          & 3.94       &   \revised{1.08}  &  \revised{0.61} & 1.21     & \textbf{0.19}              \\
               & 50,000            & 7.70          & 22.6       &   \revised{6.43}  &  \revised{3.04} & 7.84     & \textbf{0.94}              \\
               & 100,000           & 16.5          & 42.3       &   \revised{13.7}  &  \revised{6.78} & 21.7     & \textbf{2.28}              \\
               & 500,000           & 119           & 386        &   \revised{78.8}  &  \revised{30.4} & 144      & \textbf{14.6}  \\\midrule
\multirow{3}{*}{\rotatebox{90}{TC}}  
               & SF.cedge          & 439          &  296        &   \revised{443}   &   \revised{291}   &   628    & \textbf{183}    \\
               & wiki-vote         & 235          &  1372       &   \revised{391}   &   \revised{337}   &   485    & \textbf{91}  \\
               & Gnutella04       & 376          &  \timeout{}  &   \revised{441}   &   \revised{321}   &   725    & \textbf{124}  \\ 
               \bottomrule
\end{tabular}}
\end{table}

The first column of the table shows four selected Datalog queries. Galen is a Datalog version of the EL ontology \cite{kazakov2014incredible} using the ELK calculus. The three datasets used in this query are ontologies found in the Oxford Library. CSDA is a context-sensitive null-pointer analysis. The three datasets here are collected from real-world applications by the authors of a static analysis tool called Graspan \cite{Wang2017GraspanAS}. Andersen is a directed (Andersen-style) points-to flow analysis---the datasets for this query are extracted from synthesized program control-flow graphs with edge counts ranging from 10k--500k. Lastly, TC refers to a transitive closure query running on three real-world graphs from the SuiteSparse collection \cite{suitesparse}.

Columns 3, 4, 5, and 8 compare the single-core performance of all four engines. \revised{Overall, \slog{} demonstrates similar single-core performance to the Rust-based Datalog engine Nemo but is slower than RDFox.} In the Galen query tests, \slog{} consistently outperforms Nemo, though it is slower in other benchmarks. The slowest performance for \slog{} is observed with the CSDA \textsf{postgres} dataset, where it takes 1.34$\times$ longer than Nemo. On average, \slog{} is 2$\times$ slower than RDFox in single-core performance. Despite this, \slog{}, along with Nemo and RDFox, consistently outperforms VLog across all test cases. The slowdown of single-threaded \slog{} in the CSDA and Andersen queries, which involve higher memory usage than the Galen queries, is likely due to the overhead of maintaining MPI buffers, necessary for multi-core execution. However, with 12 threads, these parallel MPI facilities become worthy, making \slog{} the fastest engine overall when fully utilizing multi-core.

\revised{Columns 6 and 8 show the running times of RDFox and \slog{} with 12 threads. Leveraging data-parallelism, both engines significantly outperform the single-thread engines in all test cases when running with 12 threads. At full utilization of 12 cores, \slog{} surpasses RDFox in performance when compared at the same core counts. Compared to single-process \slog{}, multi-process execution provides an average 6$\times$ speedup, while 12-thread execution only yields a 1.5$\times$ improvement. This indicates that \slog{} has better scalability and still has  potential for further scaling with higher core counts, whereas RDFox’s scalability appears to saturate at 12 cores.}

We also benchmarked a tool that constructs on-demand why-provenance using a combination of Answer Set Programming (ASP) and SAT solvers \cite{calautti2024computing}. We attempted to run it in parallel to demand the why-provenance of all tuples in the output to simulate eager provenance; we found that the SAT solver sometimes gets stuck on lineage for certain tuples, and was not comparable with the other tools due to the inability to reuse work.

\subsection{Evaluating CFA: Slog facts vs. Souffl\'e ADTs}
\label{sec:eval:aam}
We have implemented the control-flow analysis from Figure~\ref{fig:kcfa-mcfa} in both Souffl\'e and \slog{}, and have compared their runtimes at both 8 and 64 processes on a large cloud server (with 64 physical cores) on various large synthetic benchmarks; we evaluate both $k$-CFA (exponential) and $m$-CFA (polynomial), two forms of call sensitivity.
We report our results in Table~\ref{tab:results-aam}. Each of six
distinct analysis choices is shown along the left side of the table. Along rows of
the table, we show experiments for a specific combination of analysis,
precision, and term size. We detail the total number of iterations
taken by the \slog{} backend, along with control-flow points, store
size, and runtime at both $8$ and $64$ threads for \slog{} and
Souffl\'e. Times are reported in minutes / seconds form; several runs
of Souffl\'e took under 1 second (which we mark with $<$0:01); 
\timeout{} indicates timeout (over four hours).

\input{aam-table}

Inspecting our results, we observed several broad trends. First, as
problem size increases, \slog{}'s runtime grows less-rapidly than
Souffl\'e's.
This point may be observed by inspecting runtimes for a specific
set of experiments. For example, 10-$m$-CFA with term size 200 took
\slog{} 26 seconds, while Souffl\'e's run took 56 seconds. Doubling
the term size to 400 takes 104 seconds in \slog{}, but 398 seconds in
Souffl\'e---a slowdown of $4\times$ in \slog{}, compared to a slowdown
of $7\times$ in Souffl\'e. A similar trend happens in many other
experiments, e.g., 15 minutes to over three hours for Souffl\'e
($13\times$ slowdown) vs. 2 to 4 seconds ($2\times$ slowdown) in
\slog{}'s runtime on 5-$k$-CFA. Inspecting the output of Souffl\'e's
compiled \cpp code for each experiment helped us identify the source
of the slowdown. For example, the rule for returning a value to a continuation address \texttt{\$KAddr(e,env)}, in Figure~\ref{fig:kcfa-mcfa}, must
join a return state using this address with an entry in the continuation store for this address. 

\begin{figure}[b]
\vspace{-0.35cm}
\SMALL
\begin{Verbatim}[baselinestretch=.72,commandchars=\\\{\},codes={}]
\textcolor{midorange}{// ret(av, k) :- ret(av, $KAddr(e, env)), kont_map($KAddr(e, env), k).}
\textcolor{midorange}{//        env0[0] ---^  env2[0]-^  ^--- env2[1]       env3[1] -----^} 
\end{Verbatim}
\begin{Verbatim}[baselinestretch=.72]
if(!(rel_13_delta_ret->empty()) && !(rel_18_kont_map->empty())) {
  for(const auto& env0 : *rel_13_delta_ret) {
    RamDomain const ref = env0[1];
    if (ref == 0) continue;
    const RamDomain *env1 = recordTable.unpack(ref,2);
    {
      if( (ramBitCast<RamDomain>(env1[0]) == ramBitCast<RamDomain>(RamSigned(3)))) {
	RamDomain const ref = env1[1];
	if (ref == 0) continue;
	const RamDomain *env2 = recordTable.unpack(ref,2);
	{
	  for(const auto& env3 : *rel_18_kont_map) {
\end{Verbatim}
\begin{Verbatim}[baselinestretch=.72,commandchars=\\\{\},codes={}]
	    \textcolor{midorange}{// On this line we've ommitted bitcasts and Tuple ctors:}
\end{Verbatim}
\begin{Verbatim}[baselinestretch=.72]
	    if(!(rel_19_delta_kont_map->contains({{ {{env2[0],env2[1]}}, env3[1]}}))
		&& !(rel_12_ret->contains({{env0[0], env3[1]}}))) {
\end{Verbatim}
\begin{Verbatim}[baselinestretch=.72,commandchars=\\\{\},codes={}]
	      \textcolor{midorange}{// Omitted: null checks and insertion of {{ env0[0], env3[1] }} into ret}
\end{Verbatim}
\begin{Verbatim}[baselinestretch=.72]
  }}}}}}}}}}
\end{Verbatim}
\vspace{-0.35cm}
\caption{Example \cpp code generated by Souffl\'e.}
\label{fig:souffle-cpp}
\vspace{-0.35cm}
\end{figure}
Consider the rule and its compiled \cpp code shown in Figure~\ref{fig:souffle-cpp}. Note that it uses two nested for loops to iterate over the entire \rtag{ret} relation, then iterate over the entire \rtag{kont\_map} relation, and finally check if there is a match, for every combination. In this way, Souffl\'e's lack of indices for structured values leaves it no other choice but to materialize (in time, not space) the entire Cartesian product as it checks for matches---rather than efficiently selecting tuples via an index as would be typical for top-level relations.

For a fixed problem size, we found that Souffl\'e and \slog{} both
scaled fairly well. Souffl\'e consistently performed well on small
input sizes; additional processes did not incur slowdowns, and
Souffl\'e's efficiency was generally reasonable (roughly 50\%) when
algorithmic scalability did not incur slowdowns. For example, in
3-$k$-CFA (n=$8$), Souffl\'e took 67 seconds at 8 processes, and 15
seconds at 64 processes. \slog{}'s parallelism doesn't outweigh
communication overhead on smaller problems, particularly on problems
with high iteration count and low per-iteration work. As problem size
increases, our \slog{} implementations show healthy scalability;
efficiency grows as problem size grows (e.g., 24:10 to 6:45 on
15-$m$-CFA/384, 22:46 to 7:28 on 12-$m$-CFA/800).

We found that extracting optimal efficiency
from \slog{} was facilitated by increasing per-iteration work and
avoiding long sequences of sequential work with comparatively lower
throughput; we believe this is because of the synchronization required to perform deduplication. As an example of this principle, our 10-$m$-CFA uses a
flat \rtag{ctx} fact to represent the context; a previous version used
a linked list $10$ elements deep, however this design achieved poorer
scaling efficiency due to these $10$-iteration-long sequences of work necessary to
extend the instrumentation at each call-site.
In our experiments scaling efficiency improved as polyvariance increased; e.g., improving by $2\times$ for 10-$m$-CFA, but $3.5\times$ for 15-$m$-CFA. We believe this is because of the higher per-iteration work available.

\subsection{Scaling CSPA on the Theta Supercomputer}
\label{sec:eval-cspa}
%

\begin{figure}
\begin{center}
\includegraphics[width=0.46\textwidth,clip]{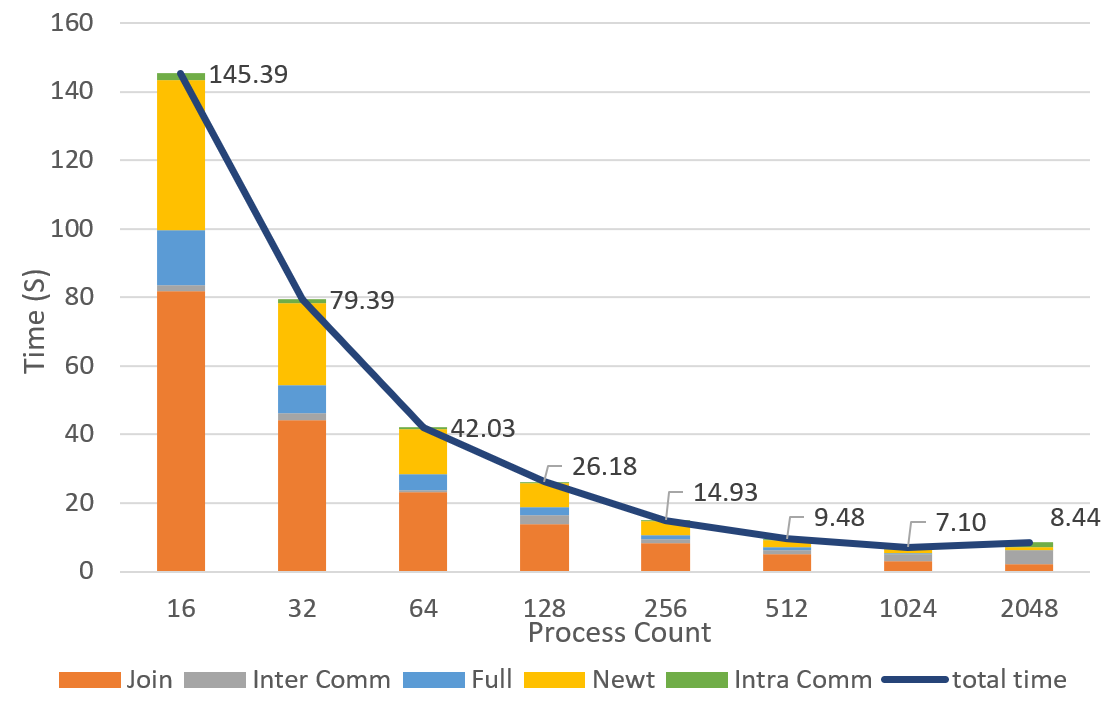}
\end{center}
\vspace{-.15in}
\caption{Scaling CSPA (of Linux) on Theta.}
\label{fig:theta}
\vspace{-.2in}
\end{figure}
The performance differences we have just observed are primarily due to the \slog{}'s asymptotic benefits in compiling rules that manipulate algebraic data. It is also reasonable to ask whether \slog{}'s data parallelism enables sufficient absolute performance in useful applications. To measure this, we transliterated (from~\cite{fan2018scaling}) a 10-rule vanilla-Datalog implementation of Context-Sensitive Points-to Analysis (CSPA) to run using \slog{}. We evaluated our implementation using several large-scale test programs from Graspan~\cite{Wang2017GraspanAS}, including \texttt{httpd}, \texttt{postgres}, and \texttt{linux} (a Linux kernel without modules). We compiled our \slog{} version of CSPA to run using MPI and ran the resulting program at a variety of scales (from 16--2,048 processes) on the Theta Supercomputer~\cite{parker2017early}
at the Argonne Leadership Computing Facility (ALCF) of the Argonne
National Laboratory. Theta is a Cray machine with a peak performance of
$11.69$ petaflops. It is based on the second-generation Intel Xeon Phi
processor and is made up of 281,088 compute cores. Theta has $843.264$ TiB of DDR4 RAM, $70.272$ TiB of MCDRAM, $10$ PiB of storage, and a Dragonfly network topology.

Figure~\ref{fig:theta} shows the result of our strong-scaling runs for CSPA of Linux (using Graspan's data). Overall, we achieve near-perfect scalability up to around 512 processes where we observe a performance improvement commensurate to change in scale. After 512 processes, scalability starts to decline owing to an increase in data movement costs and workload starvation. This trend is typical of HPC applications, which typically have a range of processes for which we observe optimal scalability. While we could not run Souffl\'e on Theta, we observed roughly similar runtimes (around 150s for both \slog{} and Souffl\'e) for this experiment at 16 threads on a large Unified Memory Access (UMA) server. We see this result as a promising indicator of the potential for \slog{} to achieve high throughput, but it is worth noting that our data set (from Graspan) achieves context sensitivity via method cloning and thus provides a large amount of available parallel work. In our future work, we aim to use \slog{} to develop rich, context-sensitive analyses of fully-featured languages.

\section{Related Work and Limitations}
\label{sec:related}

We now categorize several recent threads of related-work apropos our efforts. Along the way, we highlight limitations of our current implementation of \slog{} and discuss some plans for future work.

\revised{
\paragraph*{$\text{Datalog}^\exists$ and The Chase}
Our \dls{} is a restriction of Datalog$^{\exists}$, a well-known extension of Datalog, which allows arbitrary existential existential quantification in the heads of rules~\cite{abiteboul1995foundations}. Datalog$^{\exists}$ is particularly popular in the field of knowledge representation and reasoning (KRR),  where it enables tractable query answering over ontologies~\cite{baget2011rules}. As a result, many Datalog-based reasoners include this extension. However, existential queries are computationally expensive to compute and can also lead to undecidable queries. To address this, constrained versions of Datalog, such as those in the Datalog$^{\pm}$ family \cite{cali2009general}, are used to ensure decidability. These dialects employ various constraints, like those in the Graal reasoner~\cite{baget2015graal} and the Shy system in the i-DLV~\cite{leone2012efficiently}. 
        }

\revised{Existentially quantified conjunctive queries create tuple generating dependencies (TGDs) between the head and body clauses of a Datalog$^{\exists}$ rule. To resolve these dependencies, variants of the well-known database algorithm, the Chase~\cite{maier1979testing}, are used. One complete version is the \textit{oblivious chase}~\cite{bourhis2016guarded}, which explores all possible choices for the existentially quantified meta variables. However, this approach leads to significant non-termination issues, making it impractical for mainstream Datalog$^{\exists}$ implementations. Instead, these engines often adopt algorithms with stronger termination guarantees, such as the \textit{restricted chase}~\cite{carral2017restricted} and the \textit{parsimonious chase}~\cite{leone2012efficiently}.}

\revised{\paragraph*{Semirings and provenance}
The highly influential work of provenance semirings extends Datalog with $K$-relations, wherein every Datalog fact is labeled with a value from a semiring~\cite{Green:07}. \dls{} is orthogonal to provenance semirings. In general, wile \dls{} does allow annotating tuples with arbitrary structured values, \dls{} does not equate tuples modulo the semiring laws, instead equating facts via their Skolem term. Thus, while \dls{} does allow encoding provenance semirings (\emph{e}.\emph{g}., $\cdot(+(...),...)$), the need to materialize distinct Skolem terms (for identical semiring values) leads to encoding overhead. However, unlike provenance semirings, \dls{} does allow observing the tuple's identity: while provenance semirings forbid introspecting upon tuple tags in the Datalog program, \dls{} enables matching on structured facts to drive computation.}

Aside from provenance semirings, there has been significant interest in the application of provenance for a variety of applications, including data warehousing~\cite{Cui:2000}, debugging Datalog~\cite{Zhao:2020}, explainable AI~\cite{eriksson1985neat,ferrand2005explanations,chasing-sets}, and program analysis~\cite{cheney2009provenance}. 
Provenance has seen particular recent interest in explainable AI; our comparison systems Rulewerk and Nemo \cite{elhalawati2022existential,ivliev2023nemo} are representatives of this work, supporting why-provenance via an existential operator. Recent research leverages Answer Set Programming (ASP) and SAT solvers for why-provenance~\cite{calautti2024computing}. However, all the above approaches take an on-demand approach to computing why provenance, deeming eager materialization as unnecessary or overly expensive. By contrast, \dls{} focuses on ubiquitous, eager materialization of tree-shaped facts; our high-performance, data-parallel RA kernels (\S~\ref{sec:mpi}) enable representing provenance in a compact, distributed fashion.
\paragraph*{Equality in Datalog}
\revised{Equality-generating dependencies (EGDs) are an extension of Datalog that allow the equality operator (=) to appear in the head clauses of rules. This extension has become favored in Datalog due to its ability to facilitate new realms of application, such as financial data analysis~\cite{baldazzi2023reasoning} and program equality saturation~\cite{Willsey:2021}. One way to enable EGD reasoning is by storing Datalog relations in specialized union-find-like data structures called equality graphs (e-graphs).
Of particular note is Egglog~\cite{Yihong:2022}, which allows programming with equality sets backed by a lazy-building e-graph implementation called egg. Though techniques like lazy-building can be used to accelerate computation, full EGD chasing is still expensive. Thus, some Datalog engines, such as Vadalog, choose to use a stratified EGD semantics called \textit{harmless EGD}~\cite{bellomarini2022exploiting}, which avoids the values generated in EGD ruless from being further used to trigger the activation of other rules.}
We view \slog{} as complementary to Egglog, with similar but diverging goals and substantive differences: \slog{} does not support equality sets, though it is possible to materialize an equivalence relation (either lazily or eagerly) for bounded observations. Additionally, \slog{}'s compilation to high-performance, data-parallel kernels differentiates it from Egglog, which instead employs lazy rebuilding. Our subjective experience comparing \slog{} with Egglog is that \slog{} is faster in applications where materialization-overhead is not a bottleneck. 
We plan to study the synthesis of \slog{} and e-graphs as future work.


\paragraph*{Incremental Datalog}

A significant amount of recent interest focuses on incremental Datalog, extending Datalog's semi-na\"ive evaluation strategy to interesting domains~\cite{fixing-incremental-computation,mcsherry2013differential,dbsp,murray2013naiad}, especially focusing on generalizing Datalog's semantics in ways in which are incremental by design. Relevant systems include timely dataflow~\cite{murray2013naiad}, RDFox~\cite{rdfox}, and DBSP~\cite{dbsp}. Compared to these systems, \dls{} does not support negation or an inverse operator for elements of $\mathcal{V}$ and $\mathcal{S}$, instead preferring to be obviously-monotonic by construction. While \dls{} and \slog{} do use semi-na\"ive evaluation, they do not expose the engine's data structures in a stream-based manner. We leave to further study the connection between \slog{} with differential dataflow; we expect it is possible to compile \dls{} to DBSP, but note that our current backend is built on MPI and has the potential to leverage technologies such as InfiniBand.

\paragraph*{Distributed Datalog}

There have been significant efforts to scale
Datalog-like languages to large clusters of machines. For example, RDFox~\cite{rdfox}, 
BigDatalog~\cite{shkapsky2016big}, Distributed SociaLite~\cite{seo2013socialite},
Myria~\cite{Halperin:2014}, and Radlog~\cite{gu2019rasql} all run on
Apache Spark clusters. \slog{} differs from these systems
in two primary ways. First, compared to \slog{}'s MPI-based
implementation, Apache Spark's framework-imposed overhead is
increasingly understood to be a bottleneck in scalable data analytics
applications, with several authors noting order-of-magnitude
improvements when switching from Spark to
MPI~\cite{SparkBadMPIGood,Kumar:2017,Anderson:2017}. Second,
none of the aforementioned systems support first-class facts. We elide detailed comparison against these systems due to space; we found \slog{} was faster and more scalable than Radlog, BigDatalog, and similar systems in our limited explorations.

\section{Conclusion}

\revised{We presented \dls{}, a Datalog with first-class facts. In \dls{}, facts are identified by a symbolic representation in the form of a Skolem term. This methodology extends Datalog with the ability to introspect upon, compute with, and construct new fact identities. We rigorously define a semantics of \dls{}, presenting its semantics as a variant of the restricted chase which materializes Skolem terms, essentially giving facts an algebraic identity. The fact that rule heads cannot be unified with the body forbids the construction of cyclic facts, and thus yields a semi-decidable semantics amenable to bottom-up implementation via a simple extension of modern-generation Datalog engines.
Our higher-level language \ldls{} offers an ergonomic syntax, enabling a more direct transliteration of functional programs and natural-deduction-style rules into \dls{}. We defined \ldls{} via a syntax-directed translation into \dls{}, and showed its model-theoretic and fixpoint-based semantics, formally the equivalence of these semantics in Isabelle/HOL.}

\revised{In our implementation, we generalize \dls{} to full-fledged engine, \slog{}, in which we have implemented a wide breadth of applications, including various forms of provenance (\S\ref{sec:eval-why}) and structural abstract interpretation (\S\ref{sec:aam}). \slog{} exploits the uniqueness properties of \dls{} to ensure a massively-parallel implementation which materializes first-class facts using a communication-avoiding strategy which yields highly-scalable implementations in practice, compounding the algorithmic benefits of \dls{}. Our experiments speak to the promise of our approach: \slog{} beats all other comparison engines in our benchmarks when run at sufficient scale, and our strong-scaling runs (context-sensitive points-to analysis of Linux, \S\ref{sec:eval-cspa}) show satisfactory scalability up to 1k cores. \slog{} is available at:}

\begin{center}
\url{https://github.com/harp-lab/slog-lang1}
\end{center}

\section*{Acknowledgements}
This work was funded in part by NSF PPoSS large grants CCF-2316159 and CCF-2316157. 
This material is based upon work supported by the Defense Advanced Research Projects Agency (DARPA) under Contract No. N66001-21-C-4023. Any opinions, findings and conclusions or recommendations expressed in this material are those of the author(s) and do not necessarily reflect the views of DARPA.

\onecolumn
\begin{multicols}{2}
\bibliographystyle{ACM-Reference-Format}


\bibliography{main}
\end{multicols}

\end{document}

%% file: figure-mcfa.tex
\begin{figure}[t]
\vspace{-0.45cm}
{\small
\begin{Verbatim}[baselinestretch=.75,commandchars=\\\{\}]
\comm{// Eval states}
\rtag{ret}(v,k) :- \rtag{eval}(\rtag{$Ref}(x),env,k,_),
    \rtag{env_map}(x,env,a), \rtag{store}(a,v).
\rtag{ret}(\rtag{$Clo}(\rtag{$Lam}(x,body),env),k) :-
    \rtag{eval}(\rtag{$Lam}(x,body),env,k,_). 
\rtag{eval}(ef,env,\rtag{$ArK}(ea,env,\rtag{$App}(ef,ea),c,k),c) :-
    \rtag{eval}(\rtag{$App}(ef,ea),env,k,c).
\comm{// Ret states}
\rtag{eval}(ea,env,\rtag{$FnK}(vf,call,c,k),c) :-
    \rtag{ret}(vf,\rtag{$ArK}(ea,env,call,c,k)).
\rtag{apply}(call,vf,va,k,c) :- \rtag{ret}(va,\rtag{$FnK}(vf,call,c,k)).
\rtag{ret}(v,k) :- \rtag{ret}(v,\rtag{$KAddr}(e,env)),\rtag{kont_map}(\rtag{$KAddr}(e,env),k).
\rtag{program_ret}(v) :- \rtag{ret}(v, $Halt()).
\comm{// Apply states}
\rtag{eval}(body,\rtag{$UpdateEnv}(x,
    \rtag{$Address}(x,[call,[hist0,[hist1,nil]]]),env),
    \rtag{$KAddr}(body,\rtag{$UpdateEnv}(x,
        \rtag{$Address}(x,[call,[hist0,[hist1,nil]]]),env)),
                        [call,[hist0,[hist1,nil]]]),
\rtag{kont_map}(\rtag{$KAddr}(body,
    \rtag{$UpdateEnv}(x,\rtag{$Address}(x,
        [call,[hist0,[hist1,nil]]]),env)),k),
\rtag{store}(\rtag{$Address}(x,[call,[hist0,[hist1,nil]]]),va),
\rtag{env_update}(env,\rtag{$UpdateEnv}(x,
    \rtag{$Address}(x,[call,[hist0,[hist1,nil]]]),env)) :-
    \rtag{apply}(call,\rtag{$Clo}(\rtag{$Lam}(x,body),env),va,k,[hist0,[hist1,_]]).
\comm{// Propagate free vars}
\rtag{env_map}(x,env1,a) :- \rtag{env_update}(env,env1),
     env1 = \rtag{$UpdateEnv}(x,a,env).
\rtag{env_map}(x,env1,a) :- \rtag{env_update}(env,env1),
     \rtag{env_map}(x,env,a).
\end{Verbatim}
}
\caption{\revised{A global-store $m$-CFA---evaluated in Table~\ref{tab:results-aam}.}}
\label{fig:kcfa-mcfa}
\vspace{-0.45cm}
\end{figure}


%% file: aam-table.tex
{\footnotesize
\newcolumntype{C}[1]{>{\centering}m{#1}}
 \begin{table}
 \centering
\caption{Control-Flow Analysis: Slog vs. Souffl\'e (ADTs)}\label{tab:results-aam}
 \begin{tabular}{|m{.2cm}C{.35cm}C{.52cm}C{.7cm}C{.6cm}cccc | m{.1cm}C{.5cm}C{.75cm}C{.6cm}C{.65cm}cccc}
   \toprule 
 & \multirow{2}{*}{Size}&\multirow{2}{*}{Iters}& \multirow{2}{*}{Cf. Pts}& \multirow{2}{*}{Sto. Sz.}& \multicolumn{2}{c}{$8$ Processes} & \multicolumn{2}{c|}{$64$ Processes} \\
 & & & & & \multicolumn{1}{c|}{Slog} & \multicolumn{1}{c||}{Souffl\'e}& \multicolumn{1}{c|}{Slog}& \multicolumn{1}{c|}{Souffl\'e}\\
 \midrule 
\midrule
 \parbox[t]{2mm}{\multirow{6}{*}{\rotatebox[origin=c]{90}{3-$k$-CFA}}}
 & 8 & 1,193 & 98.1k & 23.4k & \textbf{0:01} & 1:07 & 0:02 & 00:15\\
 & 9 & 1,312 & 371.0k & 79.9k & \textbf{0:02} & 14:47 & 0:03 & 02:56 \\
 & 10 & 1,431 & 1.44M & 291.3k & 0:06 & \timeout{} & \textbf{0:05} & 45:49 \\
 & 11 & 1,550 & 5.68M & 1.11M & 0:27 & \timeout{} & \textbf{0:16} & \timeout{} \\
 & 12 & 1,669 & 22.5M & 4.32M & 2:14 & \timeout{} & \textbf{1:07} & \timeout{} \\
 & 13 & 1,788 & 89.8M & 17.0M & 12:17 & \timeout{} & \textbf{5:08} & \timeout{} \\
\midrule 
 \parbox[t]{2mm}{\multirow{6}{*}{\rotatebox[origin=c]{90}{4-$k$-CFA}}}
 & 9 & 1,363 & 311.8k & 65.4k & \textbf{0:01} & 14:38 & 0:03 & 02:08 \\
 & 10 & 1,482 & 1.20M & 229.6k & \textbf{0:05} & \timeout{} & 0:05 & 40:30 \\
 & 11 & 1,601 & 4.69M & 853.5k & 0:20 & \timeout{} & \textbf{0:13} & \timeout{} \\
 & 12 & 1,720 & 18.6M & 3.28M & 1:40 & \timeout{} & \textbf{0:53} & \timeout{} \\
 & 13 & 1,839 & 73.8M & 12.9M & 8:44 & \timeout{} & \textbf{3:58} & \timeout{} \\
 & 14 & 1,958 & 294.4M & 50.9M & 1:00:53 & \timeout{} & \textbf{35:46} & \timeout{} \\

\midrule 
 \parbox[t]{2mm}{\multirow{6}{*}{\rotatebox[origin=c]{90}{5-$k$-CFA}}}
 & 9 & 1,429 & 203.7k & 50.7k & \textbf{0:02} & 05:30 & 0:03 & 1:15 \\
 & 10 & 1,548 & 756.9k & 167.3k & \textbf{0:04} & 65:20 & 0:04 & 15:08 \\
 & 11 & 1,667 & 2.91M  & 597.5k & 0:13 & \timeout{} &  \textbf{0:08} & 3:16:06 \\
 & 12 & 1,786 & 11.4M & 2.25M & 0:56 &  \timeout{}& \textbf{0:27} & \timeout{} \\
 & 13 & 1,905 & 45.2M & 8.69M & 4:38 &  \timeout{}& \textbf{2:00} & \timeout{} \\
 & 14 & 2,024 & 179.9M & 34.2M & 25:14 & \timeout{}& \textbf{9:58} & \timeout{}\\
\midrule 
 \parbox[t]{2mm}{\multirow{6}{*}{\rotatebox[origin=c]{90}{10-$m$-CFA}}}
 & 50 & 6,120 & 21.0k & 656.8k & 0:02  & 0:02 & 00:10 & \textbf{0:01} \\
 & 100 & 11,670 & 42.9k & 2.78M & 0:07 & 0:09 & 00:20 & \textbf{0:04} \\
 & 200 & 22,770 & 86.6k & 11.4M & 0:26 & 0:56 &  00:42 & \textbf{0:23} \\
 & 400 & 44,970 & 174.0k & 46.4M  & 1:44 & 6:26 & \textbf{01:38} & 1:56 \\
 & 800 & 89,370 & 348.8k & 187.0M & 7:35 & 45:22 &  \textbf{04:21} & 9:33 \\
 & 1600 & 178,170 & 698.4k & 750.1M & 32:56 & \timeout{} & \textbf{14:36} & 1:02:35 \\
\midrule 
 \parbox[t]{2mm}{\multirow{6}{*}{\rotatebox[origin=c]{90}{12-$m$-CFA}}}
 & 25  & 3,559  & 17.1k  & 385.3k    & 0:01 & \textbf{<0:01} & 0:06 & <0:01 \\
 & 50  & 6,434  & 36.3k  & 1.89M  & 0:04 & \textbf{0:03} &  0:11 & 0:03 \\
 & 100 & 12,184 & 74.6k  & 8.28M  & 0:16 & 00:24 & 0:23 & \textbf{0:10} \\
 & 200 & 23,684 & 151.3k & 34.7M & 01:10 & 02:37 & \textbf{0:53} & 0:55 \\
 & 400 & 46,684 & 304.7k & 141.9M & 05:04 & 18:39 & \textbf{2:23} & 4:12 \\
 & 800 & 92,684 & 611.5k  & 573.8M & 22:46 & 2:38:22 & \textbf{7:28} & 24:58 \\
\midrule 
 \parbox[t]{2mm}{\multirow{6}{*}{\rotatebox[origin=c]{90}{15-$m$-CFA}}}
 & 12  & 2,211  & 14.5k  & 136.7k     &  0:01 & \textbf{<0:01} & 0:04 & <0:01 \\
 & 24  & 3,591  & 35.9k  & 1.44M   &  0:03 & 0:02 & 0:06 & \textbf{0:01} \\
 & 48  & 6,351  & 78.6k  & 8.29M   &  0:14 & 0:15 & 0:14 &  \textbf{0:07} \\
 & 96  & 11,871 & 164.2k & 38.9M  &  01:08 & 01:41 & \textbf{0:36} & \textbf{0:36} \\
 & 192 & 22,911 & 335.4k & 168.0M &  05:15 & 12:10 & \textbf{1:49} & 2:51 \\
 & 384 & 44,991 & 678k & 697M & 24:10 & 1:32:35 & \textbf{6:45} & 16:32 \\
\bottomrule
\end{tabular}
\end{table}
}